\documentclass[%
 reprint,
superscriptaddress,
 amsmath,amssymb,
pra,
floatfix,
]{revtex4-1}
\usepackage[T1]{fontenc}

\usepackage{graphicx}
\usepackage{dcolumn}
\usepackage{bm}
\usepackage{amsmath,amssymb}
\usepackage{xcolor}
\usepackage{makecell}
\usepackage[detect-all]{siunitx}
\usepackage[normalem]{ulem}

\usepackage[colorlinks=false]{hyperref}
\usepackage{color,soul}

\newcommand{\diagdots}[3]{%
  \rotatebox{#1}{\makebox[0pt]{\makebox[#2]{\xleaders\hbox{$\cdot$\hskip#3}\hfill\kern0pt}}}%
}

\newcommand\figWidth{86}

\DeclareSIUnit\gauss{G}

\begin{document}

\preprint{PRA}

\title{Applications of maximum likelihood estimations for analyzing photon counts in few atom experiments}

\author{M.~Weyland}
\affiliation{The Dodd-Walls Centre for Photonic and Quantum Technologies, University of Otago, Dunedin, New Zealand}
\affiliation{Department of Physics, University of Otago, Dunedin, New Zealand}

\author{L.~Sanchez}
\affiliation{The Dodd-Walls Centre for Photonic and Quantum Technologies, University of Otago, Dunedin, New Zealand}
\affiliation{Department of Physics, University of Otago, Dunedin, New Zealand}

\author{P.~Ruksasakchai}
\affiliation{The Dodd-Walls Centre for Photonic and Quantum Technologies, University of Otago, Dunedin, New Zealand}
\affiliation{Department of Physics, University of Otago, Dunedin, New Zealand}

\author{M.~F.~Andersen}
\email{mikkel.andersen@otago.ac.nz}
\affiliation{The Dodd-Walls Centre for Photonic and Quantum Technologies, University of Otago, Dunedin, New Zealand}
\affiliation{Department of Physics, University of Otago, Dunedin, New Zealand}

\date{\today}

\begin{abstract}
We present a method for determining the atom number distribution of few atoms in a tight optical tweezer from their fluorescence distributions.  In the tight tweezer regime, the detection light causes rapid atom loss due to light-assisted collisions. This in turn leads to non-Poissonian and overlapping fluorescence distributions for different initial atom numbers, and commonly used threshold techniques fail. We use maximum likelihood estimation algorithms to fit model distributions that account for the atom loss. This gives accurate atom number distributions for relatively few experimental runs (about 600 is sufficient) to sample a photon number distribution. We show that the method can be extended to situations when the photon number distributions for known initial atom numbers cannot be modeled, at the cost of requiring a higher number of experimental runs.
\end{abstract}

\maketitle

\section{Introduction}
Optical tweezers have enabled studies of individual atoms and their interactions~\cite{Gruenzweig2010, Liu2018, Weyland2021, Andersen2022}. Furthermore, arrays of single atoms in optical tweezers have become a very promising platform for new quantum computers~\cite{Weitenberg2011, Graham2022} and provide an important tool for studying many-body physics~\cite{Bernien2017, Browaeys2020} and atom entanglement~\cite{Dordevic2021, Wilk2010}.
In few-atom experiments, a standard way of probing an outcome is to count the number of atoms that remain after an interaction loss event~\cite{Gruenzweig2010, Fuhrmanek2010, Reynolds2020, Weyland2021, Gruen2024, Zhang2020}. 
For instance, a common technique to determine the internal state of atoms in an optical tweezer relies on ejecting the atoms in a particular internal state and counting the number of atoms remaining~\cite{Kuhr2005}. 
Therefore, improvements of methods to determine the atom number proportions in a tight tweezer have numerous applications. 

When imaging multiple atoms in a tight optical tweezer, light-assisted collisions lead to atom loss~\cite{Gruenzweig2010}. This loss limits the amount of detection light scattered by the atoms. As a result, the photon count distributions obtained from different numbers of atoms overlap and the shapes of these distributions are generally non-Poissonian~\cite{Xu2021, Bloch2023}. Therefore, simple threshold techniques~\cite{McGovern2011} or fits with Poisson distributions~\cite{Meng2020} do not determine the atom number distributions. Nevertheless, one can fit with measured photon count distributions for known atom numbers~\cite{Sompet2019, Reynolds2020, Weyland2021}.

Here, we examine a method to determine the atom number proportions in a tight optical tweezer. It is based on measuring the fluorescence distributions for different known atom numbers in control measurements. These recorded distributions are used to determine the loss rates during atom detection and to create model fluorescence distributions. A weighted sum of these model distributions is fitted to the fluorescence distribution of an unknown atom number distribution. The fitting is done using maximum likelihood estimation algorithms. We analyze its performance when using the modeled fluorescence distributions for known atom numbers, and when using recorded distributions. We find that using the modeled distributions gives more accurate results in particular when fluorescence distributions are sampled with limited statistics. We further show that maximum likelihood algorithms work better than simple least-squares fitting, whose basic assumptions are not valid. The uncertainty in the fitted parameters are determined using a bootstrap method.   

The remainder of the paper is structured as follows: In Sec.~\ref{Sec2:PhotonCountMesurement}, we describe our experimental setup and how we acquire the photon count distributions. Section~\ref{Sec3:DistributionModel} presents a model for the fluorescence distribution of an arbitrary number of atoms in presence of atomic loss during the exposure. In Sec.~\ref{Sec4:AtomNumberDetermim}, we describe the maximum likelihood estimator used to fit the photon count distributions. The fitting performance is assessed in Sec.~\ref{Sec5:PoissModelFitPerformance}. Section~\ref{Sec6:DataFit} shows an alternative approach relying on recorded distributions instead of models for the fluorescence distribution. We then describe our error estimation in Sec.~\ref{Sec7:ErrorsRobustness} and test the fits with modified fluorescence distributions as well as check the dependence of the fluorescence distribution on the atom temperature in Sec.~\ref{Sec8:Robustness}.

\section{Measurement of Photon count distributions}\label{Sec2:PhotonCountMesurement}
We now give a brief overview of the experimental setup and sequence that we use to acquire the photon count distributions from a known number of atoms in the tweezer. More details on our experimental procedures are given in Refs.~\cite{McGovern2011,Carpentier2013,Hilliard2015,Fung2016,Reynolds2020}. 

\subsection{Preparation of a known number of atoms in the tweezer} 
\begin{figure}[h!]
\includegraphics[width=\figWidth mm]{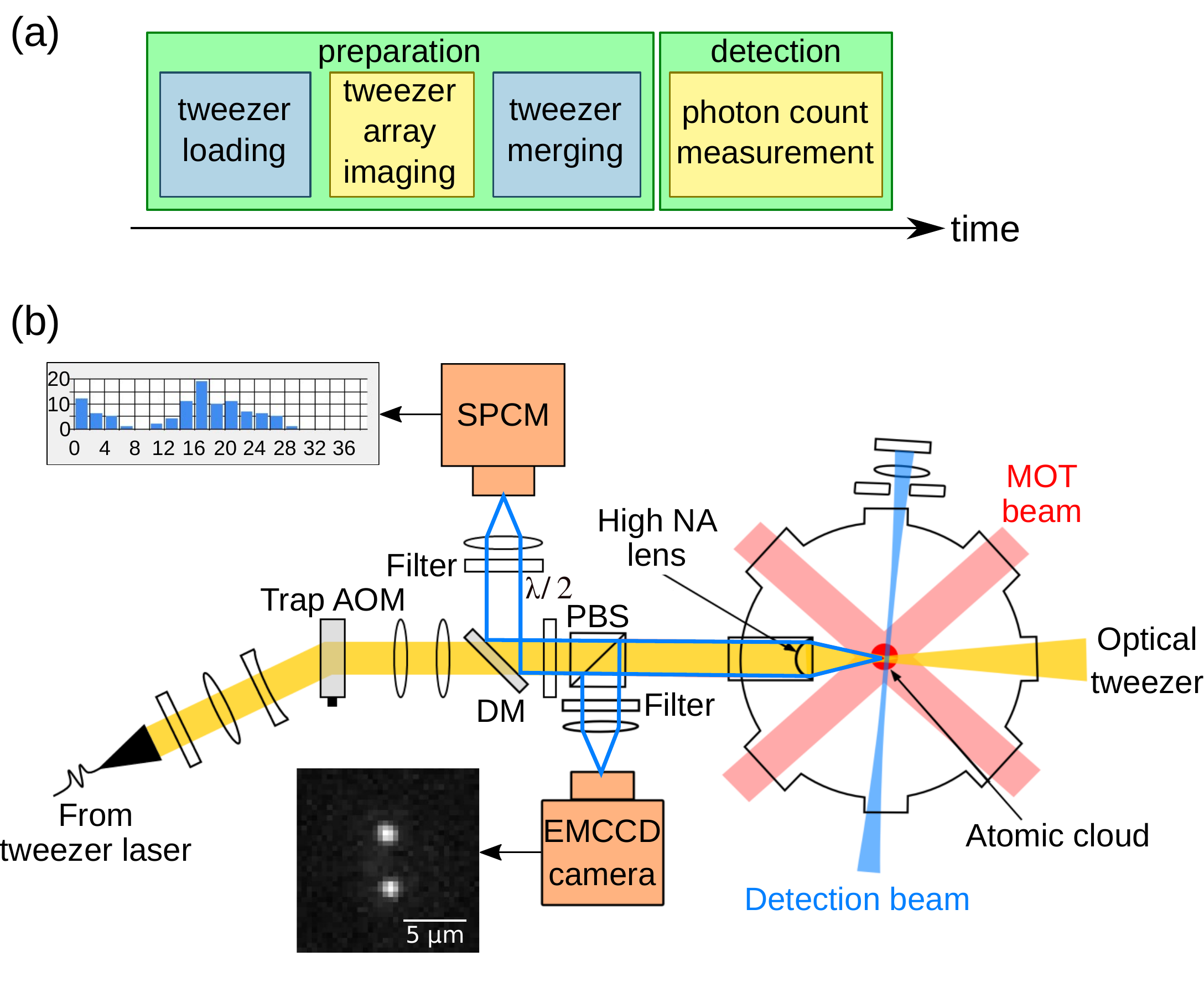}
	\caption{(a) Timeline of a typical experiment to study individual atoms. (b) Sketch of the experiment: The atoms are trapped in the center of the vacuum chamber by the optical tweezer (yellow beam). The detection beam is focused on the atoms and induces fluorescence. The MOT beams on the $D_2$ line provide cooling of the atoms during imaging. 
    The fluorescence is collected via the high NA lens, filtered from scattered light of the tweezer by a dichroic mirror and band-pass filters, and detected by the EMCCD camera and by the single photon counting module (SPCM, blue lines).}
	\label{fig:sketch}
\end{figure}

The experimental sequence to prepare a known number of atoms in the tweezer has three parts, which are shown in Fig.~\ref{fig:sketch}(a).  
During the loading stage, several $^{85}$Rb atoms are captured from a MOT into each of two or more optical tweezers separated by 5\,$\mu$m. The tweezers are produced from the same 1064\,nm laser beam deflected by an acousto optic modulator (AOM) driven by multiple RF tones. The emerging deflected beams are subsequently focused in the science chamber using a high NA lens (Fig.~\ref{fig:sketch}(b)). The depths and spacing of the tweezers are controlled by the amplitude and frequencies of the RF tones. Blue detuned light-assisted collisions on the $D_1$ line assure that each tweezer contains no more than one atom and facilitate a high single atom loading efficiency as described in Refs.~\cite{Gruenzweig2010, Carpentier2013, Fung2015}. During the tweezer array imaging stage, fluorescence generated by exposing the atoms to the retro-reflected detection beam allows us to determine how many tweezers successfully captured a single atom. The detection beam is focused to a beam waist of 190\,$\mu$m (full width at half maximum) at the position of the atoms and contains two light frequencies resonant with transitions on the $D_1$ line of the atoms. A frequency component resonant with the  $F=3 \rightarrow F'=3$ transition at 2\,$\mu$W prevents build-up of population in the $F=3$ ground state. The other component is blue detuned from the $F=2 \rightarrow F'=3$ transition at a power of 16.5\,$\mu$W. It provides laser cooling and causes the atoms to fluoresce when they are in this state. Concurrently, the six beams, that are also used for the initial loading, provide three dimensional cooling. 
A proportion of the 795\,nm fluorescence from the $D_1$ line is collected through the high NA lens and imaged onto an Andor iXon 897 EMCCD camera, with an exposure time of \SI{17}{\milli\second}. Interference filters in front of the camera prohibit the detection of light from the six cooling beams.  
Fig.~\ref{fig:sketch}(b) includes an example image of two atoms loaded into separate optical tweezers. Post-selection on such images allows us to sort the experimental realizations according to the number of atoms loaded. If the fluorescence level detected by the EMCCD for a given tweezer is above a given threshold, the tweezer contains an atom. This determination of successful loading is superior to 99.7\% of cases leading to negligible miscounts. There is no significant loss during this imaging stage as each tweezer contains no more than one atom, so there are no light-assisted collisions. Concurrently with imaging, cooling in the $k_B \times$\SI{2.7}{\milli\kelvin} deep tweezers to around \SI{15}{\micro\kelvin} assures that no atom loss occurs at this stage. Before starting the merging step, we adiabatically ramp the depth of the tweezers to \SI{1.5}{\milli\kelvin} by decreasing the amplitude of the RF tones to the AOM. The tweezers are then moved close together by sweeping the RF frequency for one of the tweezers toward the other in 20\,ms. At the end of this sweep the tweezers are \SI{0.85}{\micro\metre} apart. Since our optical tweezers have a beam waist of \SI{1.05}{\micro\metre}, their combined potential after the sweep has a single minimum~\cite{Sompet2019}.  
One of the tweezers is then turned off in 20\,ms leaving a single optical tweezer with a known number of atoms.

\subsection{Photon count measurement and histograms}
The detection stage records the photon counts from a tweezer that contains a known number of atoms, when exposed to the same detection beam configuration as during the tweezer array imaging stage.
To do so, part of the 795\,nm fluorescence photons collected by the high NA lens is sent toward a single photon counter module (SPCM), as depicted in Fig.~\ref{fig:sketch}(b).  After separating the 795\,nm light from the tweezer beam via a dichroic mirror (DM) and an additional band-pass filter, the SPCM detects around 4\% of the photons emitted by the atoms in the tweezer.  A micro-controller records the number of photons detected by the SPCM during a $T_{\text{exp}}=$\,\SI{3.5}{\milli\second} exposure time.

\begin{figure}[h!]
\includegraphics[width=\figWidth mm]{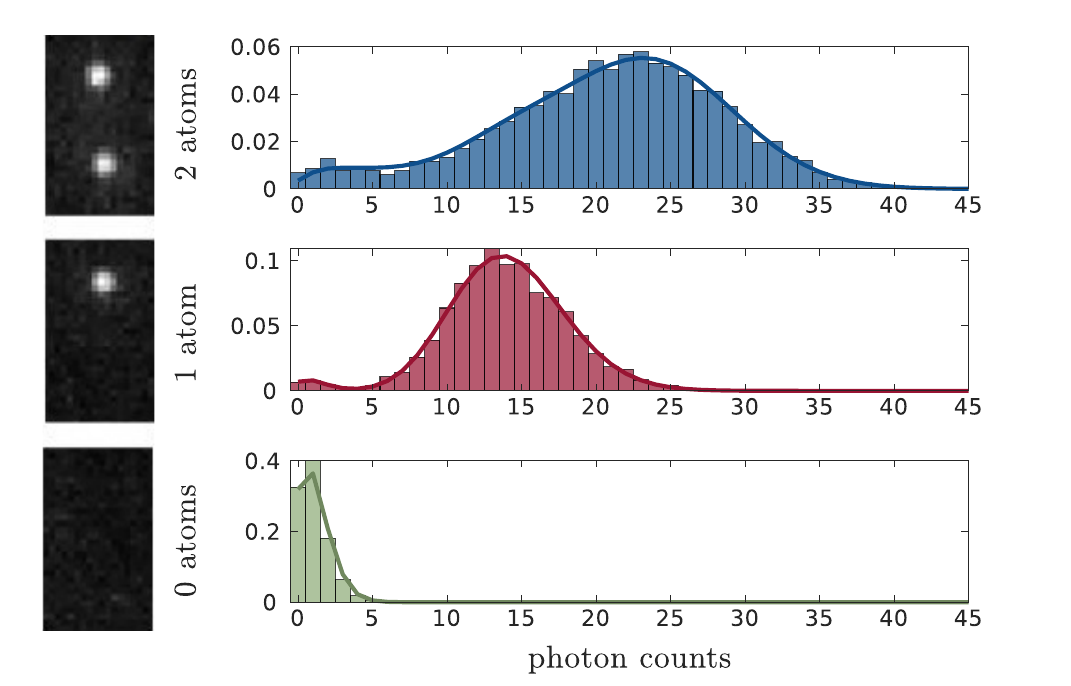} 
	\caption{Example of measured histograms created from a total of 11200 experimental realizations, representing the photon count distributions when detecting two, one and zero atoms in an optical tweezer via imaging (left side pictures). The solid curves are the theoretical fits which will be described in Sec.~\ref{sec:basis}.}
	\label{fig:histograms}
\end{figure}

The experimental sequence is repeated hundreds of times. The photon count recorded during the detection stage is sorted by the number of atoms that the tweezer contains, using the fluorescence image obtained during the preparation stage. An example of histograms showing the photon count distribution for up to two atoms present in the tweezer is displayed in Fig.~\ref{fig:histograms}.

When no atoms are present the counts follow a Poisson distribution with a mean of about 1 photon as shown by the green histogram in Fig.~\ref{fig:histograms}. The counts mainly corresponds to scattered 795\,nm detection light by the experimental apparatus, tweezer light not entirely filtered out and dark counts by the SPCM. The 1-atom photon counts in Fig.~\ref{fig:histograms} show a peaked distribution with an average photon count around 15 corresponding to the photons scattered by one atom in the tweezer. Additionally, it displays a small peak related to a contribution of the 0-atom distribution originating from the atom being lost before measuring photon counts. This single-atom loss contribution will be discussed in Sec.~\ref{sec:basis}. The 2-atom distribution peaks at around 23 photons, exhibiting a wide shape with significant overlap with the 0-atom and the 1-atom photon count histograms. Although the different photon count distributions are not separable, they still display distinct shapes.

\section{Fluorescence distribution modeling}\label{Sec3:DistributionModel}
For determining the proportion of zero-, one- or two- atoms remaining in the tweezer after an experiment, one can fit an experimental photon distribution with a weighted sum of the photon distributions from particular atom numbers. For this we define the $i-$atom normalized photon distribution $b_{i,k}(T_\mathrm{exp})$, which gives the probability for observing $k$ photons for an exposure time of $T_\mathrm{exp}$ when there are $i$ atoms present at the beginning of the exposure.

\subsection{General model}
We make two assumptions to be able to model the photon count distribution for known initial atom numbers.
First, we assume that for a duration ($\tau$) when the atom number ($i$) does not change, the photon counts recorded are Poisson distributed:
\begin{equation}
P(k;\lambda_i) = \frac{\lambda_i^k}{k!}e^{-\lambda_i}
\end{equation}
where $k$ is the number of photons and $\lambda_i=\eta_i \tau$ the photon count average and $\eta_i$ the photon detection rate with $i$ atoms present~\cite{goodman2015}. Secondly, we assume that loss events occur at fixed rates given by $\gamma_{ij}$, where $i$ is the number of atoms present and $j$ is the number of atoms left after a loss event. The combined loss event rate when $i$ atoms are present is therefore: $\Gamma_i=\sum_{j=0}^{i-1}{\gamma_{ij}}$.
 
The above assumptions allow for a recursion relation for the photon count distributions $b_{i,k}$ as defined previously, assuming that the photon distributions are known for all atom numbers smaller than $i$:

\begin{widetext}
\begin{align}
  &b_{i,k}(T)=  e^{-\Gamma_i T } P\left( k;\eta_i T \right) +\hspace{-0.5em} \int\limits_0^{T}{\hspace{-0.3em} e^{\ -\Gamma_i \tau } \sum_{j=0}^{i-1} {\gamma_{ij}}
  \sum_{\kappa=0}^k{P\left(\kappa; \eta_i \tau \right) b_{j, k-\kappa}\left( T-\tau \right)}d\tau} \label{gi}
\end{align}
\end{widetext}
The first term describes the possibility that no loss occurred, while the second accounts for one or more loss events with the first at any time $\tau$ during $T$. If there are no atoms present, no loss can occur, so $b_{0,k}(T)$ is a Poisson distribution with mean $\lambda=\eta_0 T$. Expressions for any higher atom numbers are then found recursively using Eq.~\ref{gi}.

\subsection{Application to two atom experiment}\label{model}
We now focus on the situation where the tweezer can contain up to two atoms. This situation requires the modeled photon distributions for an exposure $T=T_\mathrm{exp}$ for an initial atom number of 0, 1, and 2. 
The 0-atom distribution is a Poisson distribution of mean $\eta_0 T_{\text{exp}}$, as discussed previously. For the 1-atom case, we neglect the possibility of having the atom lost during the exposure, giving $\gamma_{10}=0$. In our experiment, the timescale for a 1-body loss process under the detection beam is at least three orders of magnitude larger than $T_{\text{exp}}$, making this a good assumption. The 1-atom photon count distribution is therefore also assumed to be a Poisson distribution, but with a mean of $\eta_1 T_{\text{exp}}$. This assumption is supported by the Poisson shape of the 1-atom control histogram (Fig.~\ref{fig:histograms}). If this assumption is not fulfilled as in~\cite{Bloch2023}, the recursion relation can be used.
When two atoms are present, light-assisted collisions can lead to atom loss during the exposure. As a result the 2-atom histogram in Fig.~\ref{fig:histograms} is not Poissonian. A collisional loss event can result in one or both atoms being lost~\cite{Gruenzweig2011}. The resulting 2-atom photon distribution obtained from Eq.~\ref{gi} is:

\begin{align}
    &b_{2,k}(T_{exp})= e^{-\Gamma_2 T_{\text{exp}}}\ P(k;\eta_2 T_{\text{exp}}) \nonumber \\
    &+ C_0\sum_{\kappa=0}^k\left( P(\kappa;\beta_0 \eta_0 T_{\text{exp}}) -P(\kappa;\beta_0 \eta_2 T_{\text{exp}}) \right)\\
    &+C_1 \sum_{\kappa=0}^k\left( P(\kappa;\beta_1 \eta_1 T_{\text{exp}}) -P(\kappa;\beta_1 \eta_2 T_{\text{exp}}) \right) \nonumber \\
&\text{with:}\quad \beta_j = 1+\frac{\Gamma_2}{\eta_2 -\eta_j} \nonumber \\
&C_j =\frac{\gamma_{2j}(\eta_2 -\eta_j)^k}{ \left[(\eta_2-\eta_j)+\Gamma_2\right]^{k+1}}e^{\frac{\Gamma_2\eta_j T_{\text{exp}}}{(\eta_2-\eta_j)}} \nonumber
\end{align}

The input parameters ($\{\eta_j\}$ and $\{\gamma_{ij}\}_{j<i}$), where $\{ ... \}$ represents the set, are found by fitting histograms such as those shown in Fig.~\ref{fig:histograms} with a Maximum Likelihood Estimation algorithm.

\section{Principle of determination of atom number distribution}\label{Sec4:AtomNumberDetermim}
Next, we present the method used to deduce the atom number proportions in our experiment, applying Maximum Likelihood Estimation (MLE) algorithms.

\subsection{Likelihood fitting algorithms}
The commonly used least-squares fitting method is generally not suitable for fitting count data. Indeed, its basic assumptions are not fulfilled, \textit{e.g.} as bin counts are non-negative, the errors cannot be symmetric~\cite{goodman2015, Coxe2009}. We therefore use the Maximum Likelihood Estimation (MLE) method to determine the most likely set of parameters $\Vec{\theta}$, that would yield the measured histogram $\{ m_k \}$.

For independent measurements, such as photon counts recorded by the SPCM, the likelihood of $\Vec{\theta}$ given $\{ m_k \}$ is: $\mathcal{L}(\Vec{\theta} | \{m_k\}) = \prod_k p\left( m_k | \Vec{\theta} \right)$, where $p\left( m_k | \Vec{\theta} \right)$ is the probability of having the count $m_k$ in bin $k$ given the parameter set $\Vec{\theta}$. 
For computational purposes, we use the logarithmic likelihood (log-likelihood) which gives the following estimator that has to be maximized to find $\Vec{\theta}$ giving $L_{\text{max}}$:
\begin{equation}
    L(\Vec{\theta},\{m_k\}) = \ln \mathcal{L}(\Vec{\theta},\{m_k\}) = \sum_{k} \ln\left( p\left( m_k | \Vec{\theta} \right)\right)
    \label{eq:max-log-likelihood}
\end{equation}

There are different methods to evaluate the probability $p$ present in the likelihood function $L$ in Eq.~\ref{eq:max-log-likelihood} depending on the algorithm used. 

\subsection{Determination of the input parameters for $\{b_{i,k}\}$}\label{sec:basis}
The next step is to determine the input parameters ($\{\eta_j\}$ and $\{\gamma_{ij}\}_{j<i}$)) of the modeled functions $\{b_{i,k}\}$ from Section~\ref{Sec3:DistributionModel}. These are determined from experimentally acquired photon distributions for known initial atom numbers such as those displayed in Fig.~\ref{fig:histograms}. To do so we use the following procedure in which all fits are done with an MLE algorithm~\footnote{We use the MLE function of MATLAB for this.}. We start by fitting the 0-atom photon count distribution described as a Poisson law $b_{0,k}=P(k; \eta_0T_{\text{exp}})$ to the recorded 0-atom photon count distribution (lower panel in Fig.~\ref{fig:histograms}). This determines the parameter $\eta_0$. 
As mentioned earlier, the 1-atom histogram (middle panel of Fig.~\ref{fig:histograms}) shows two parts: the fluorescence of a single atom in the tweezer as a Poisson distribution peaked at $\eta_1 T_{\text{exp}}$ and a small contribution of $b_{0,k}$ corresponding to the absence of atoms at the start of photon counting. This difference of the number of atoms between imaging and detection is interpreted as a probability $\ell$ to lose the atom ahead of detection. 
This parameter gathers single atom losses proceeding from collisions with the background gas, from the merge process and false 1-atom detection with the EMCCD camera. Therefore, we fit the 1-atom histogram by a weighted sum of two Poisson distributions such as $\ell P(k; \eta_0T_{\text{exp}}) + (1-\ell) P(k; \eta_1 T_{\text{exp}})$, where $\eta_0$ is already fixed. 
The loss $\ell$ and the rate $\eta_1$ are the adjustable parameters~\cite{Raikov1937}.  This results in $b_{1,k}(T_{\text{exp}}) = P(k; \eta_1 T_{\text{exp}})$.
Finally, we determine the rate parameters ($\eta_2$ and $\{\gamma_{2j}\}_{j<2}$) of the $b_{2,k}$ model by fitting the 2-atom histogram with a weighted sum of the three model distributions while accounting for the single-atom losses as: $\ell^2 b_{0,k} + 2\ell(1-\ell) b_{1,k} +(1-\ell)^2 b_{2,k}$. Again, previously determined quantities ($\ell$, $b_{0,k}$, and $b_{1,k}$) are kept fixed while $\eta_2$ and $\{\gamma_{2j}\}_{j<2}$ in $b_{2,k}$ are adjusted. 
This weighted sum used to calculate $b_{2,k}$ assumes that there is no additional loss induced by the presence of two atoms compared to when only one atom is present. Any additional two-body loss will manifest itself as a higher proportion of zero atoms, leading to more photon counts close to zero. Such a feature of the distribution cannot be reproduced by adjusting the rate parameters. Therefore, additional loss processes would be detected when the fit of the two-atom distribution fails. In our experiment, we observe no additional loss processes. 
The results of the fits are displayed in solid lines in Fig.~\ref{fig:histograms}. We see the models match the measured photon number distributions well. 
The $\{b_{i,k}\}$ set is then fully determined for fitting to the photon distribution from an unknown atom number distribution in the tweezer. 

\subsection{Atom-number distributions determined with Poisson-loglikelihood}
\begin{figure}[h!]
\includegraphics[width=\figWidth mm]{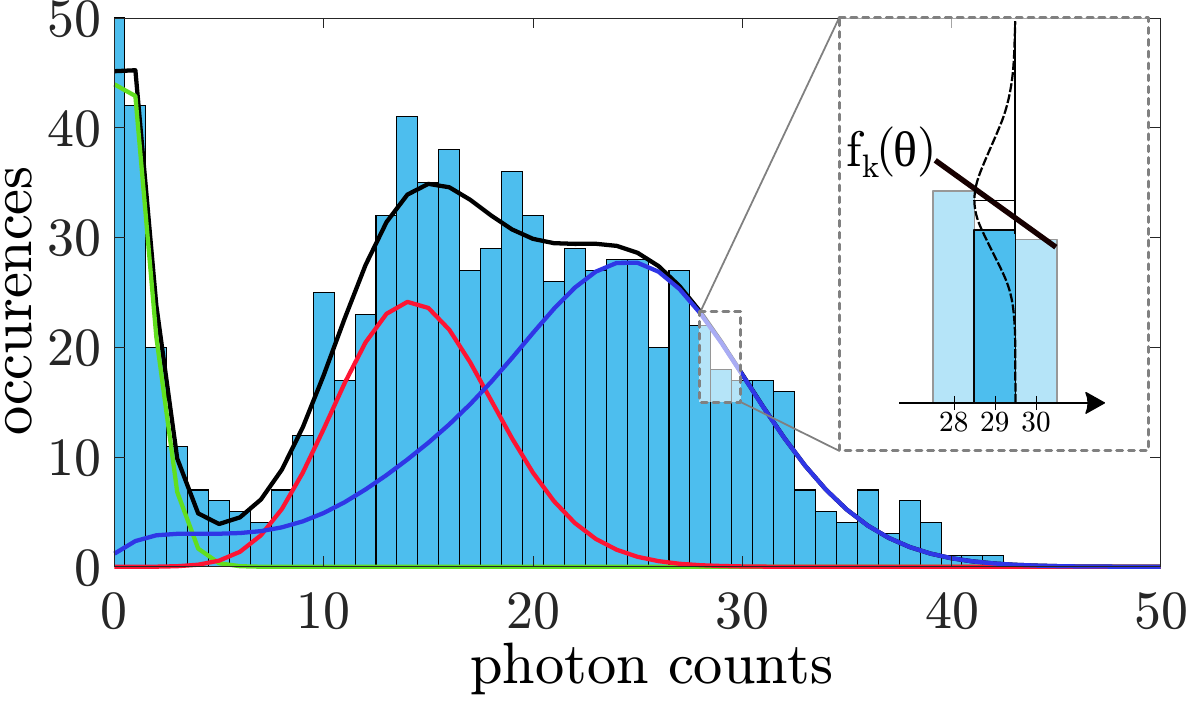}
	\caption{Example of a fitting result: the test histogram $m_k$ proceeds from the compilation of $800$ experimental realizations. The overall analytical model $f_k(\Vec{\theta})$ fitted to the test histogram is shown in black. The weighed components $N_d a_0 b_{0,k}$, $N_d a_1 b_{1,k}$, $N_d a_2 b_{2,k}$ are displayed as a line in green, red and blue, respectively. The insert focuses on a particular bin $k=29$: The thick black line is the value of the model distribution evaluated in this bin ($f_{29}(\Vec{\theta})$) and the dashed line indicate the Poisson distribution with the mean equal to $f_{29}(\Vec{\theta})$.
 }
	\label{fig:fitEx}
\end{figure}

We now determine the proportions of zero, one and two atoms given a set of  $N_d$ photon counts from an experimental realization in which the atom number distribution is, in principle, unknown. $m_k$ is the measured occurrence for $k$ detected photons and it is fitted with a model distribution for the expected mean for bin $k$ that is a weighted sum of $b_{i,k}$ rescaled to the number of measured events $N_d$:
 \begin{equation}
 f_k(\Vec{\theta}) = N_d \sum_i a_i b_{i,k},
    \label{eq:fit-basis}
\end{equation} 
where $\Vec{\theta} =(a_0,a_1, a_2)$ is the fitting parameter with $a_i$ the weighting of $i$ atom(s) in the tweezer. To evaluate the likelihood estimator $L(\Vec{\theta},\{m_k\})$, the simplest and most common method relies on choosing the probability function to follow a Poisson law. This is illustrated in the inset of Fig.~\ref{fig:fitEx}. The thin black line shows the Poisson distribution with the mean $f_k(\theta)$, which is evaluated at the experimental occurrences, shown by the height of the blue bar. As each bin count comes from discrete and independent events, the content of bin $k$ is Poisson distributed such that $p\left( m_k | \Vec{\theta} \right) =\mathrm{P}(m_k;f_k(\Vec{\theta}))$~\cite{Barlow1989, taylor}.
 The Poisson log-likelihood method is valid when any potential uncertainty in $b_{i,k}$ used in Eq.~\ref{eq:fit-basis} can be neglected. We assume this to be the case for our modeled fluorescence distributions. In practice, fitting the atom number distribution for up to two atoms requires only two free parameters, $a_0$ and $a_1$, with the constraint of  $a_2 + a_1 + a_0 =1$.
This limited parameter space allows us to perform an exhaustive parameter search. To ensure finding the global optimum we use a 3-stage grid search algorithm with a final grid size of 0.025\%~\footnote{While more sophisticated fitting algorithms would be able to perform this fit, we chose this simple approach as it can also serve as a robust method when directly fitting to experimentally measured control distributions with no functional model, as described in Sec.~\ref{Sec6:DataFit}.}.
The result of such a fit is illustrated in Fig.~\ref{fig:fitEx}, where the modeled functions are plotted against a fitted experimental photon count distribution.

\section{Poisson likelihood fitting performance}\label{Sec5:PoissModelFitPerformance}

To test the  method we fit Eq.~\ref{eq:fit-basis} to histograms produced from known atom numbers proportions. The fitting performance is assessed as a function of two parameters: $N_d$ which is the number of data points in the histogram proceeding from a mixture of different atom numbers, and $N_c$ which is the number of data points in the recorded histograms of known atom numbers such as those in Fig.~\ref{fig:histograms} from which $b_{i,k}$ is determined.

\subsection{Test distributions}
We run the experimental sequence described in Sec.~\ref{Sec2:PhotonCountMesurement} in blocks of 200 realizations. Each of these blocks is randomly assigned to either the \textit{test} pool or the \textit{control} pool.
The \textit{test} and \textit{control} pools created this way have a total size of 6400 data points each, consisting of 56\% 2-atom realizations, 38\% 1-atom realizations and 6\% 0-atom realizations as a result of the experimental single atom loading efficiency of 75\%.

 For a test, we first construct a set of 3 control histograms similar to Fig.~\ref{fig:histograms} from a total of $N_c$ data points taken from the \textit{control} pool by random sample with replacement. We keep the percentage of 2-, 1- and 0-atom realizations as in the original pool. The subsequent histograms are fitted with the modeled photon count distributions as described in Sec.~\ref{sec:basis}, to determine the $\{b_{i,k}\}$ basis. Next, we create a mixed data set with $N_d$ data points with a known atom number proportions $(\bar a_0, \bar a_1, \bar a_2)$ by random sample with replacement from the \textit{test} pool making the test histogram $\{m_k\}$. Finally, we used the Poisson-loglikelihood algorithm to retrieve the atom number proportions from which the test histogram was made and assess how close the fit result $\Vec{\theta} = (a_0, a_1, a_2)$ is to the constructed mix $(\bar{a}_0, \bar{a}_1, \bar{a}_2)$.

\subsection{Goodness-of-fit}
\label{Sec5B:GOF}
We evaluate the performance of the fit while varying the number of data points in the control histograms as well as in the test histogram. 
To measure the discrepancy of the fit, we calculate the Root Mean Square Error (RMSE) from the errors $a_i-\bar{a}_i$ between the fit results and the expected ones, with $i$ the atom number.
We then obtain a representative deviation $\Delta$ by averaging the RMSE over the whole parameter space $\{\Vec{\theta}\}$ using 10\% steps and performing 100 resamples of the photon counts for every $\{\Vec{\theta}\}$:

\begin{equation}
    \Delta = \left< \sqrt{\frac{1}{n+1}\sum_{i=0}^n  \left(a_i-\bar{a}_i\right)^2} \right>_{\Vec{\theta},r} .
    \label{eq:Delta}
\end{equation}
Here, $n$ is the maximum number of atoms, $1<r<100$ is the resampling index and the $\langle ...\rangle_{\theta,r}$ represent the averaging over all $\theta$ and $r$. The closer $\Delta$ is to zero, the more accurate is the fit result.
Furthermore, we calculate the 68\%-confidence interval (CI) of the proportion of $i$ atoms from the distribution of their errors and use the size of the CI to visualize the expected spread in the fit results.

\begin{figure}
\includegraphics[width=\figWidth mm]{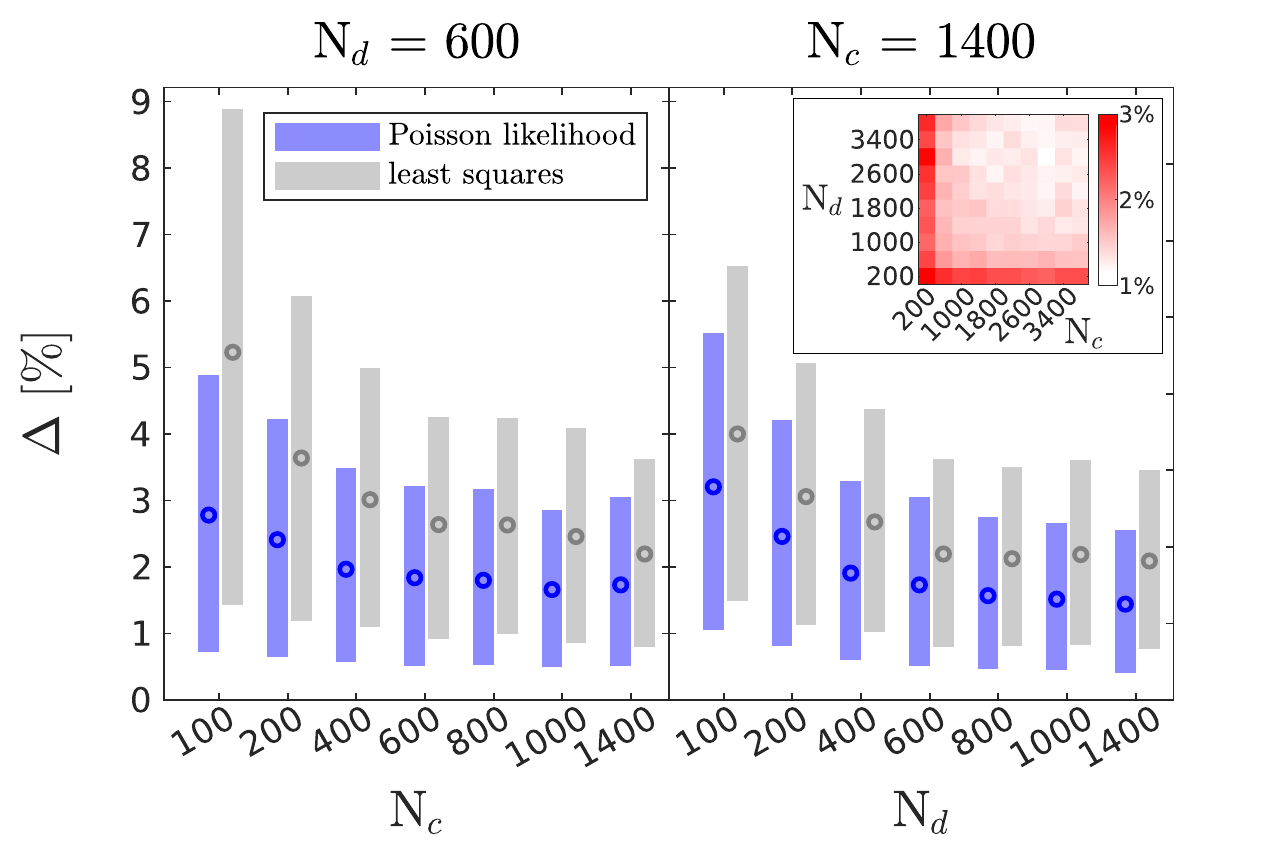}
	\caption{Deviation $\Delta$ (open circles) when fitting a photon histogram containing $N_d$ data points using the modeled photon count distributions. Maximum Poisson likelihood is in blue, and a least-squares fit is in gray. The bars show the one $\sigma$ confidence interval as described in the text. Left: Deviation with $N_d=600$ while changing the amount of control data $N_c$. Right: Deviation with $N_c=1400$ while changing the amount of test data $N_d$. The inset shows $\Delta$ when varying both $N_d$ and $N_c$ while maximizing the Poisson likelihood.}
	\label{fig:change_N_BG_2methods}
\end{figure}

Figure~\ref{fig:change_N_BG_2methods} shows the fit deviations $\Delta$ of the maximum likelihood fit and their confidence interval (blue) for different quantities of control and test events. 
As the number of control events increases the fits are more accurate: the deviations decrease and so do their confidence intervals. 
For both the control dataset as well as the test dataset, the decrease in $\Delta$ diminishes when increasing the amount of data in the varied dataset (either $N_c$ or $N_d$). When one dataset is significantly larger than the other one, $\Delta$ is dominated by the fluctuations in the smaller dataset. For instance, in Fig.~\ref{fig:change_N_BG_2methods} (left), $\Delta$ reaches its minimum value around $N_c = 800$, as $N_d = 600$. Instead of increasing the amount of data in one of the sets, it is advantageous to increase both datasets to achieve higher fit accuracy, as can be seen in the inset in Fig.~\ref{fig:change_N_BG_2methods}. Over the whole tested range of $N_c$ and $N_d$, the least-squares fit (shown in gray) shows higher deviations, caused by its lack of suitability.

\begin{figure}
\includegraphics[width=\figWidth mm]{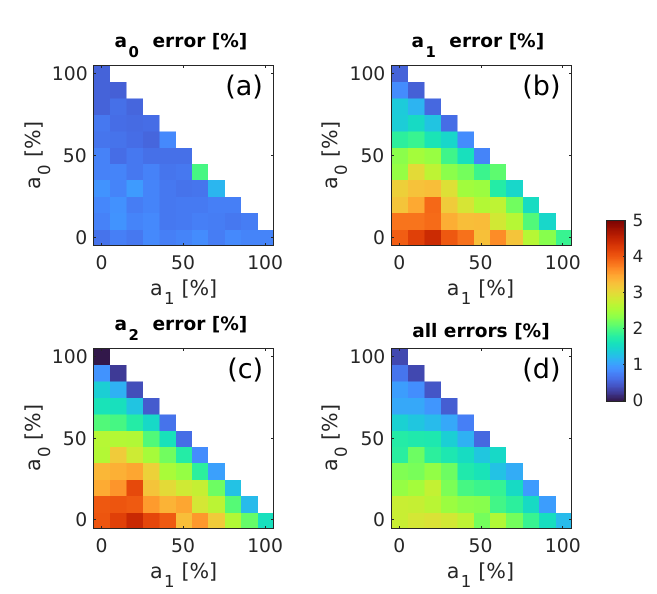}
	\caption{(a)-(c) Root Mean Square Errors of the different fitted atom proportions $a_i$ as a function of input atom proportions. The whole parameter space $\vec\theta=(a_0,a_1,a_2)$ is covered in 10\% steps. (d) Combination of all errors calculated as $\Delta$ but before averaging over $\vec\theta$. These results were obtained with $N_d=600$ and $N_c=1400$, using the Poisson maximum likelihood fit.}
	\label{fig:triangle}
\end{figure}

Next, we investigate how the fit error varies depending on the atom number proportions. Since we just showed that maximizing the Poisson likelihood gives superior results over a least-square fit, we use only results of the Poisson maximum likelihood fit for this analysis.
We calculate the RMSE of all atom proportions $a_i$, where the mean is taken over the resamples $r$. The results are shown as a function of $a_0$ and $a_1$ in Fig.~\ref{fig:triangle}, since $a_2 =1-a_1-a_0$. The further a point is on the bottom left of the plot, the higher the proportion of $a_2$.
We make two observations:  First, we observe a low error for $a_0$ (shown in Fig.~\ref{fig:triangle}(a)\,), independent of the chosen $\vec\theta$. 
Second, the errors in $a_1$ and $a_2$ (shown in Fig.~\ref{fig:triangle}(b)-(c)\,) are largest when the $a_2$ is high. Both these observations are consistent with errors that are larger for stronger overlapping distributions. Importantly, the fit works over the whole parameter space, including extreme cases where some proportions are equal to zero. Fig.~\ref{fig:triangle}(d) shows the combination of all errors, calculated like $\Delta$, but without averaging over $\vec\theta$. While the errors vary with proportions, there are no regions where they reach values that render the fit unusable. Hence, $\Delta$ obtained by averaging over $\theta$ serves as a good measure of performance. 

\section{Event-based fitting}\label{Sec6:DataFit}
Situations may occur where a model for the i-atom photon distribution is not known. In such situations it is possible to directly use measured distributions $b_{i,k}$ as was done for example in Refs.~\cite{Sompet2013, Sompet2019, Reynolds2020}. 
We proceed to investigate how such an approach compares to using the model from Sec.~\ref{Sec3:DistributionModel}. In the following, the fit using the measured histograms will be named ``event-based fit'' whereas the one using model distributions will be referred to as ``model-based fit'' for distinction. 

\subsection{Effective likelihood and its implementation}
When model distributions cannot be derived, the normalized histograms proceeding from experimental acquisitions of photon counts with a known atom number (Fig.~\ref{fig:histograms}) can provide a fitting basis. These histograms inherently contain cross-talks if there is single atom loss prior to detection (parameterized as $\ell$) but if it can be determined the fitted atom proportions $\{a_i\}$ can be post-corrected. More details are given in the Appendix. 

The Poisson likelihood method is valid when any potential uncertainty in $f_k$ in Eq.~\ref{eq:fit-basis} can be neglected. It can therefore yield unaccounted errors when the amount of control events is limited~\cite{Glusenkamp2020}. There are numerous ways to account for statistical fluctuations in $b_{i,k}$, by adjusting the likelihood calculations~\cite{Bohm2012,Bohm2014,Chirkin2013,Glusenkamp2018}. 
Here, we use the \emph{effective} likelihood method introduced in Ref.~\cite{Arguelles2019} alongside the standard Poisson likelihood and compare their performance. The effective likelihood accounts for statistical fluctuations in the \textit{control} histograms in addition to fluctuations in the experimental histogram to be fitted.
Instead of assuming that$f_k(\Vec{\theta})$ for bin $k$ represents the mean of Poisson distributed events, the effective likelihood treats the value $f_k(\Vec{\theta})$ as a possible outcome for a random draw from a Poisson distribution with an unknown average, leading to the following likelihood~\cite{Arguelles2019}:

\begin{equation}
    \mathcal{L}_{\mathrm{Eff},k}(\Vec{\theta} | m_k) = \int_0^\infty P(m_k;\lambda) \mathcal{G}\left(\lambda;\frac{f_k^2(\Vec{\theta})}{\sigma_k^2(\Vec{\theta})}+1,\frac{f_k(\Vec{\theta})}{\sigma_k^2(\Vec{\theta})}\right) d\lambda
\end{equation}

where $\mathcal{G}$ is the Gamma distribution and

\begin{align}
    \sigma_k^2(\Vec{\theta}) &= \sum_i b_{i,k} \left(a_i N_d\right)^2 / N_{c,i}
\end{align}
where $N_{c,i}$ is the number of control events with $i$ atoms loaded in the tweezer, given the total number of control events $ N_c =\sum_i N_{c,i}$. 
$\sigma_k(\Vec{\theta})$ is a measure of the statistical uncertainty of the expected counts in bin $k$ proceeding from the limited number of events $N_{c,i}$ in the control histogram $b_{i,k}$. 
The log-likelihood to be maximized during the fitting process is then: 
\begin{equation}
    L(\theta, \{ m_k \}) = \sum_k \ln{\mathcal{L}_{\mathrm{Eff},k}(\Vec{\theta} | m_k)} .
    \label{eq:Leff_to_maximize}
\end{equation}

\subsection{Fitting performance}
\begin{figure}
 \includegraphics[width=\figWidth mm]{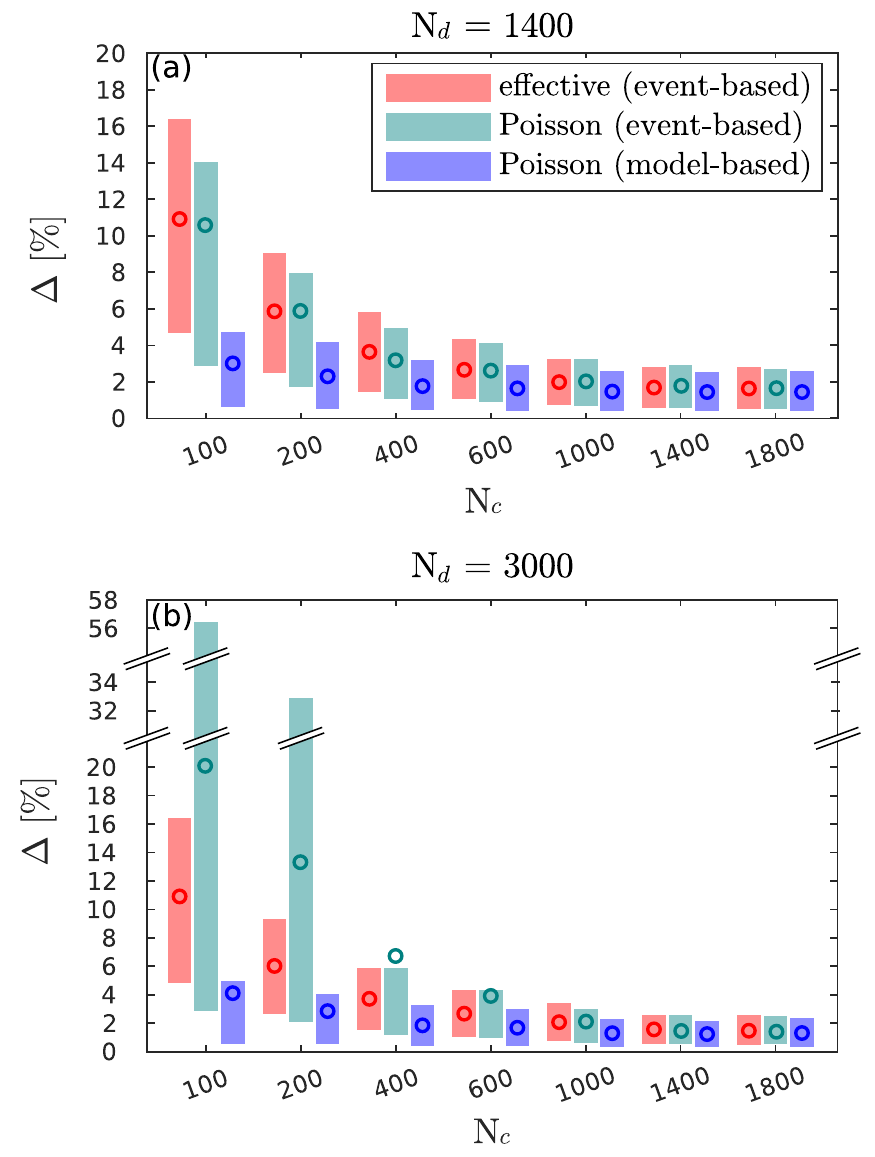}
	\caption{Deviation $\Delta$ (open circles) and its CI (colored bars) when fitting a test histogram of $N_d$ photon counts for different number of control events $N_c$ used for the distribution fit basis. The event-based fits are using the \emph{effective} likelihood (red) and the Poisson likelihood (green). The model-based fit results is displayed for comparison (blue). (a) $N_d=1400$ test photon counts.  (b) $N_d =3000$ test photon counts. }
	\label{fig:data_fits}
\end{figure}

Like in Sec.~\ref{Sec5:PoissModelFitPerformance}, we calculate the deviation $\Delta$ and its confidence interval for different amounts of $N_d$ and $N_c$, but using the event-based fits.
In Fig.~\ref{fig:data_fits} we compare the result of three different fitting methods: The event-based effective likelihood fit maximizes the likelihood in eq.~\ref{eq:Leff_to_maximize}, where the $b_{i,k}$ are the measured distributions. The event-based Poisson likelihood fit maximizes eq.~\ref{eq:max-log-likelihood} with $p\left( m_k | \Vec{\theta} \right) =\mathrm{P}(m_k;f_k(\Vec{\theta}))$ the Poisson likelihood, where the $b_{i,k}$ are again measured distributions. Lastly, for comparison we show the fit using the model-based Poisson likelihood, which was previously used in Sec.~\ref{Sec5:PoissModelFitPerformance}. 

Figure~\ref{fig:data_fits}(a) shows the deviation $\Delta$ as a function of the size of the control data set $N_c$ with the size of the test data set fixed at $N_d=1400$. 
The accuracy of the event-based fit improves as the number of $N_c$ events increases, but the effective likelihood does not perform better than the Poisson likelihood for the sizes of the data sets considered here. The deviations from model-based Poisson maximum-likelihood fit are also shown for comparison. The model-based fit performs better for low $N_c$, but for $N_c >600$, the event-based fits' performance approaches that of the model-based fit. It is not surprising that model-based fit performs better at low $N_c$ as it is more robust to statistical noise in the control sets.

Figure~\ref{fig:data_fits}(b) shows the deviations when $N_d$ is increased to 3000, thereby having less statistical fluctuations in the data histogram. The event-based Poisson likelihood fit now fails for low $N_c$. Inspection of the parameter-space reveals that the fit first fails in certain regions of $\theta$-space~\footnote{When the fit fails in a small region of $\theta$-space, the average error increases without a significant impact on the $1\sigma$ confidence interval. As a result the average error starts being located towards the upped edge or even outside of the confidence interval.}. Failure of the Poisson likelihood fit for low $N_c$ is expected as we are trying to fit under-sampled distributions to a well-defined test histogram, and the Poisson likelihood fit does not account for statistical fluctuations in the $b_{i,k}$. On the other hand, the effective likelihood gives sensible results although the deviation increases significantly for low $N_c$. A balance between the number of events for constructing the control and experimental histograms is preferable. The better performance of the effective likelihood method is anticipated as it is expected to be able to deal with the situation when fluctuations in the control histograms are not negligible. Again, model-based fitting outperforms event based fitting, so this should be the preferred option when possible. 

\section{Bootstrap error estimation}\label{Sec7:ErrorsRobustness}
There are uncertainties associated with the atom number distributions returned by the fitting procedure. We now account for the statistical fluctuations inherent in count data by estimating the errors on a fitted sample. Furthermore, we investigate how potential variability in physical parameters in the experiment may affect the results. 

Calculating the errors associated with the log-likelihood fit is difficult when there is no obvious analytical formula for the propagation of errors.
However, as the experimental data consists of a random sample from an underlying unknown distribution, it is well suited to error estimation via bootstrapping.
 To estimate the errors of the fit of a particular experimental histogram $\{m_k\}$ containing $N_d$ events, we generate 100 bootstrap photon count lists of the size $N_d$ by randomly sampling with replacement from the photon count list. The succeeding bootstrap histograms are then fitted using the same MLE algorithm as the original data. The 1$\sigma$ bounds of the bootstrap results are used as the confidence intervals of the obtained best fit parameters $(a_0, a_1, a_2)$.

For this method to be appropriate, the measured data set of $N_d$ photon count events need to be representative of the real underlying distribution. For independent and identically distributed observed values, this assumption is verified as soon as the data sample is large enough~\cite{Chernick}. 
Therefore, we compare the bootstrap confidence intervals relying on resampling only from one specific set of size $N_d$ to the confidence intervals of fit results proceeding from resample sets from the entire \textit{test} and \textit{control} data pools, as used in the previous section. 

\begin{figure}[h!]
\includegraphics[width=\figWidth mm]{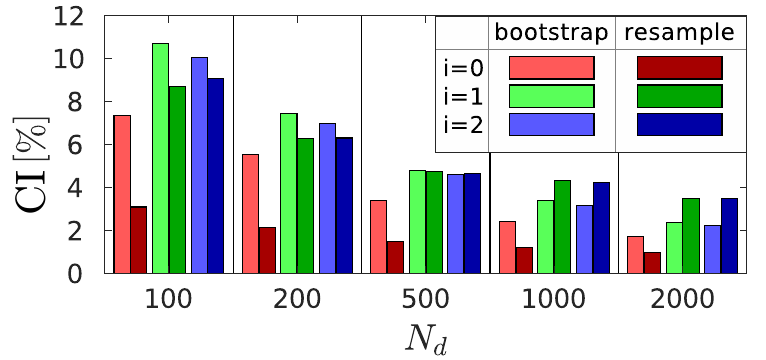}
	\caption{Evolution of the confidence intervals of the atomic proportions $\{a_i\}$ with the number of test events $N_d$. The control events are kept constant at $N_c=2000$. The bootstrap CIs (light bars) for model-based fits are shown alongside with the CIs obtained from 100 resamples of the \textit{test} and \textit{control} event sets (dark bars). }
	\label{fig:CI_comparison}
\end{figure}

Figure~\ref{fig:CI_comparison} presents a comparison of the size of the CIs calculated from the resampled set fits and the ones from bootstrapping on a particular set. The data is from model-based fits, but event-based fits show similar results. 
We see that bootstrapping captures the CIs well for the parameters $(a_1, a_2)$ for the range of $N_d$ tested. There is a tendency to that the bootstrap method overestimates the error in $a_0$, but it remains a reasonable error estimate. 

\section{Change in photon count histogram}\label{Sec8:Robustness}

The shapes of the photon count distributions are crucial for our fitting method. In this section we test how deliberately changing the detection light alters the photon count distributions and fitting performance.
Afterwards, we test how changing the temperature of the atoms at the start of the photon count measurement impacts the distributions. 

\subsection{Changing detection light}\label{Subsec:changing_hist}

We have shown how to obtain the atom number distribution from a photon count histogram and that the bootstrap technique is able to capture the uncertainty due to the statistical fluctuations. We will now investigate this result for photon count distributions with different amounts of overlap. 
To test the determination of the number distribution under different experimental conditions, we deliberately changed the detection beam during the detection stage. 

\begin{figure}[h!]
 \vspace{0.5cm}
 \includegraphics[width=\figWidth mm]{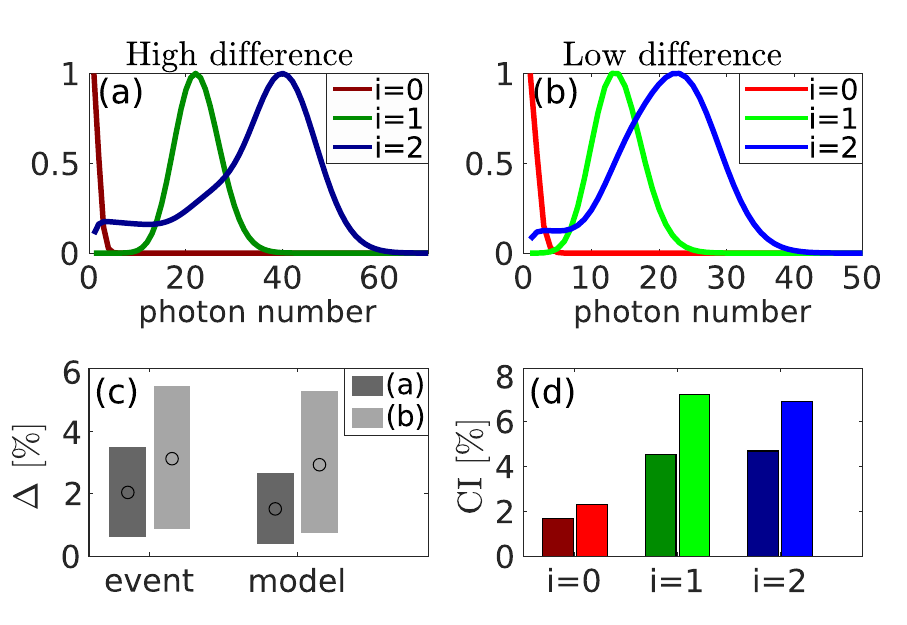}
	\caption{Comparison of fitting performances using two different sets of control distributions. The distributions shown in (a) exhibit a high difference between the one- and two-atom photon count histograms, while those in (b) show a low difference. The deviation $\Delta$ shown in (c) is calculated from fitted results using the Poisson likelihood method, with $N_d=400$ test events and $N_c=1000$ control events. The size of the confidence intervals calculated via resamples of both test and control data is shown in (d). In both (c) and (d) the dark graphs correspond to the distributions in (a) while the light graphs correspond to the distributions in (b).}
	\label{fig:BG_comp}
\end{figure}

A comparison of two different modeled photon distributions is shown in Fig.~\ref{fig:BG_comp}(a) and (b). In (a) the detuning of the $F=2 \rightarrow F'=3$ component of the detection light was 13\,MHz, while in (b) it was 19\,MHz and the $F=3 \rightarrow F'=3$ component was turned off. We still detect fluorescence from the $F=2 \rightarrow F'=3$ component, as the beams providing three-dimensional cooling prevent complete transfer of population to the $F=3$ ground state. The adjustments cause the one-atom and two-atom photon count distributions in Fig.~\ref{fig:BG_comp}(b) to show a significantly lower difference than in (a). 

Figure~\ref{fig:BG_comp}(c) displays the deviation $\Delta$ for both distributions. The deviation $\Delta$ increases and its confidence interval increases with stronger overlapping distributions (lighter bars). Both model-based and event-based fitting profit from decreased overlap between the histograms. The confidence intervals in Fig.~\ref{fig:BG_comp}(d) shows the same behavior: The stronger overlapping distributions cause larger confidence intervals in particular for the one and two atom proportions.  
However, even the strongly overlapping distributions in Fig.~\ref{fig:BG_comp}(b) give meaningful results, leading to the conclusion that it is desirable but not essential to find experimental parameters that decrease the overlap of the photon count distributions.

\subsection{Atom temperature before photon count measurement}\label{Subsec:changing_temp}

When performing an experiment with two atoms in a tweezer, the studied process can change the temperature of the atoms. 
As a result, the atoms detected in control measurements may have a different temperature than those detected after an experiment. Therefore, we need to verify that the photon count distributions are not impacted by the atom temperature. 

\begin{figure}[h!]
 \vspace{0.5cm}
 \includegraphics[width=\figWidth mm]{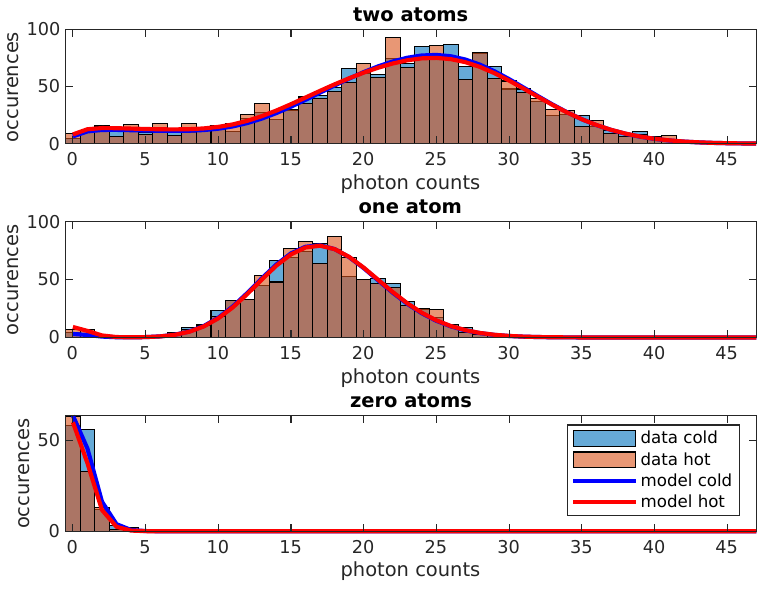}
	\caption{Comparison of photon count distributions for zero, one and two atoms obtained from atoms at 27\,$\mu$K (blue bars) and 68\,$\mu$K (orange bars). The model fits are shown in blue and red lines.}
	\label{fig:temp_comp}
\end{figure}

We collect two sets of photon count distributions. In one set, the atoms have a temperature of 27\,$\mu$K (at a tweezer depth of $k_B \times$\SI{2.7}{\milli\kelvin}) after merging the tweezers. In the other set the temperature is 68\,$\mu$K after the tweezer merge, due to different laser cooling parameters. 
For both the ``cold'' and the ``hot'' preparations, we collect 2400 events. This is higher than the number of control events required for good fitting performance in Sec.~\ref{Sec5:PoissModelFitPerformance} and Sec.~\ref{Sec6:DataFit} to make this measurement more sensitive to possible changes in the photon count distributions. 
The resulting photon count distributions as well as their fitted model distributions for zero, one and two atoms can be seen in Fig.~\ref{fig:temp_comp}. We observe a very good agreement between the two sets of measurements, with a minor change in single atom loss $\ell$, from 0.8\% for the ``cold'' sample to 2.0\% for the ``hot'' sample. 
To obtain a measure for the agreement between these distributions, we perform a $\chi^2$-test, using the model fitted to the ``cold'' distributions as the expected values and the histograms of the ``hot'' atoms as the observed values, and \textit{vice versa}. 
These tests result in reduced-$\chi^2$ values of 1.2 ($p=0.11$) and 0.9 ($p=0.65$), respectively, where $p$ is the probability to find a larger reduced-$\chi^2$ value, assuming our observations follow the expected distributions. These values are consistent with the hypothesis that both distributions are the same at a 5\% significance level~\cite{taylor}.

This observed insensitivity of the photon count distributions to the atom temperature is not a universal feature. It is at least partially enforced by our experimental design. Our detection light also acts as cooling light. This configuration prevents atom loss due to heating during the detection stage and rapidly rethermalizes the atom to a new equilibrium temperature independent of the original one. When such an implementation is not possible, it may be necessary to measure the temperature after the experiment and ensure that the control measurements are obtained at the same temperature.

\section*{Summary and Outlook}\label{Sec8:Summary}
We investigated a method to determine the atom number distribution present in a tight optical tweezer from to its fluorescence distribution. It works in the regime when atom loss causes the photon distributions for different atom numbers to overlap. We derived model distributions for the photon counts from known initial atom numbers when losses occur during the exposure time. Fits of these to photon count distributions using Maximum Likelihood Estimation algorithms allow us to determine unknown atom number distributions. A bootstrap method gives good estimation of the errors in the fitted parameters. Using about 2000 experimental runs for determining the model distribution parameters and for sampling the photon count distribution for unknown atom number distributions gives errors on the order few percentage points in the fitted atom number proportions. 

If model distributions cannot be found, it is still possible to determine unknown atom number distributions using event-based fitting, which utilizes measured photon count distributions from known atom numbers to fit to the photon count distributions from unknown atom number distributions. However, event-based fitting generally requires better statistics in photon count distributions to reach an accuracy similar to that of the model-based fitting. If the photon count distribution from the unknown atom number distribution has good statistics, then an effective likelihood method performs better than the simpler Poisson likelihood for event-based fitting. The effective likelihood accounts for statistical fluctuations in the photon distributions for known atom numbers. 

The analysis focused on photon count distributions with up to two $^{85}$Rb atoms trapped in our optical tweezer. 
However, the improvements in the atom number determination presented here for up to two atoms can be extended to experiments with three or more atoms~\cite{Reynolds2020,Weyland2021}, where more rudimentary analysis methods struggle to determine atom numbers. It can serve as an important tool for studying the emergence of many-particle behavior from few particle systems. Furthermore, recent experiments with other elements observed loss during single atom imaging~\cite{Bloch2023,Gruen2024}. These can also use the proposed model to improve atom detection and measure loss rates.

\section*{Acknowledgment}
This work was supported in part by Quantum Technologies Aotearoa, a research programme of Te Whai Ao – the Dodd Walls Centre, funded by the New Zealand Ministry of Business Innovation and Employment through International Science Partnerships, contract number UOO2347. Additionally, it was supported by the Marsden Fund Council from Government funding, administered by the Royal Society of New Zealand (Contract No. UOO1835).

\appendix

\section{Corrections for atom loss}
We have seen that single atom loss before detection cause contamination of the measured $\{ b_{i,k} \}$. This may lead to systematic errors in the atom number distributions determined by event-based fitting. However, if the single atom loss parameter $\ell$ can be determined, the atom number distributions can be corrected for it, and the systematic error eliminated. In our comparison of event based fitting to model based fitting we have corrected the event based fitting for the effect of $\ell$, to avoid our comparison to be skewed by the systematic error in the event based fitting.    
To determine the single atom loss parameter $\ell$, we fit the measured 1-atom distribution with a sum of two Poisson distributions. As an example, in Fig.~\ref{fig:histograms} of the main text, this loss probability is equal to 2.9\%. This method of determining $\ell$ would naturally not work if the photon count distribution for one atom was not Poissonian. 
To correct for finite $\ell$, the atom number distribution determined by fitting (here named $\Tilde{\theta}=(\Tilde{a}_0,\Tilde{a}_1,\Tilde{a}_2)$  to distinguish them from the final results) are post-corrected to give the final tweezer populations $\theta=(a_0, a_1, a_2)$: 

\begin{align}
    \label{eq:population_correction1}
    {a_0} &= \Tilde{a}_0 + \ell \Tilde{a}_1 + \ell^2 \Tilde{a}_2 \\
    \label{eq:population_correction2}
    {a_1} &= (1-\ell) \Tilde{a}_1 + 2 \ell (1-\ell) \Tilde{a}_2 \\
    \label{eq:population_correction3}
    {a_2} &= (1-\ell)^2 \Tilde{a}_2.
\end{align}

To cover the whole parameter space of ${a_i} \in [0,1]$ for the final probabilities, the parameter space in which the best fit is found gets expanded beyond the $[0,1]$ interval depending on the measured single atom loss. This allows fitting of data with less atom loss than observed in the control data.
From Eqs.~\ref{eq:population_correction1}-\ref{eq:population_correction3} we obtain that $ \Tilde{a}_0 \in [-\ell,1]$, $ \Tilde{a}_1 \in [-2\ell,1/(1-\ell)]$ and $ \Tilde{a}_2 \in [0,1/(1-\ell)^2]$. We use these intervals to find the best fits.

\bibliography{mybibliography}

\providecommand{\noopsort}[1]{}\providecommand{\singleletter}[1]{#1}%
\begin{thebibliography}{41}%
\makeatletter
\providecommand \@ifxundefined [1]{%
 \@ifx{#1\undefined}
}%
\providecommand \@ifnum [1]{%
 \ifnum #1\expandafter \@firstoftwo
 \else \expandafter \@secondoftwo
 \fi
}%
\providecommand \@ifx [1]{%
 \ifx #1\expandafter \@firstoftwo
 \else \expandafter \@secondoftwo
 \fi
}%
\providecommand \natexlab [1]{#1}%
\providecommand \enquote  [1]{``#1''}%
\providecommand \bibnamefont  [1]{#1}%
\providecommand \bibfnamefont [1]{#1}%
\providecommand \citenamefont [1]{#1}%
\providecommand \href@noop [0]{\@secondoftwo}%
\providecommand \href [0]{\begingroup \@sanitize@url \@href}%
\providecommand \@href[1]{\@@startlink{#1}\@@href}%
\providecommand \@@href[1]{\endgroup#1\@@endlink}%
\providecommand \@sanitize@url [0]{\catcode `\\12\catcode `\$12\catcode
  `\&12\catcode `\#12\catcode `\^12\catcode `\_12\catcode `\%12\relax}%
\providecommand \@@startlink[1]{}%
\providecommand \@@endlink[0]{}%
\providecommand \url  [0]{\begingroup\@sanitize@url \@url }%
\providecommand \@url [1]{\endgroup\@href {#1}{\urlprefix }}%
\providecommand \urlprefix  [0]{URL }%
\providecommand \Eprint [0]{\href }%
\providecommand \doibase [0]{http://dx.doi.org/}%
\providecommand \selectlanguage [0]{\@gobble}%
\providecommand \bibinfo  [0]{\@secondoftwo}%
\providecommand \bibfield  [0]{\@secondoftwo}%
\providecommand \translation [1]{[#1]}%
\providecommand \BibitemOpen [0]{}%
\providecommand \bibitemStop [0]{}%
\providecommand \bibitemNoStop [0]{.\EOS\space}%
\providecommand \EOS [0]{\spacefactor3000\relax}%
\providecommand \BibitemShut  [1]{\csname bibitem#1\endcsname}%
\let\auto@bib@innerbib\@empty
\bibitem [{\citenamefont {Grünzweig}\ \emph {et~al.}(2010)\citenamefont
  {Grünzweig}, \citenamefont {Hillard}, \citenamefont {McGovern},\ and\
  \citenamefont {Andersen}}]{Gruenzweig2010}%
  \BibitemOpen
  \bibfield  {author} {\bibinfo {author} {\bibfnamefont {T.}~\bibnamefont
  {Grünzweig}}, \bibinfo {author} {\bibfnamefont {A.}~\bibnamefont {Hillard}},
  \bibinfo {author} {\bibfnamefont {M.}~\bibnamefont {McGovern}}, \ and\
  \bibinfo {author} {\bibfnamefont {M.~F.}\ \bibnamefont {Andersen}},\
  }\href@noop {} {\bibfield  {journal} {\bibinfo  {journal} {Nat. Phys.}\
  }\textbf {\bibinfo {volume} {6}},\ \bibinfo {pages} {951} (\bibinfo {year}
  {2010})}\BibitemShut {NoStop}%
\bibitem [{\citenamefont {Liu}\ \emph {et~al.}(2018)\citenamefont {Liu},
  \citenamefont {Hood}, \citenamefont {Yu}, \citenamefont {Zhang},
  \citenamefont {Hutzler}, \citenamefont {Rosenband},\ and\ \citenamefont
  {Ni}}]{Liu2018}%
  \BibitemOpen
  \bibfield  {author} {\bibinfo {author} {\bibfnamefont {L.~R.}\ \bibnamefont
  {Liu}}, \bibinfo {author} {\bibfnamefont {J.~D.}\ \bibnamefont {Hood}},
  \bibinfo {author} {\bibfnamefont {Y.}~\bibnamefont {Yu}}, \bibinfo {author}
  {\bibfnamefont {J.~T.}\ \bibnamefont {Zhang}}, \bibinfo {author}
  {\bibfnamefont {N.~R.}\ \bibnamefont {Hutzler}}, \bibinfo {author}
  {\bibfnamefont {T.}~\bibnamefont {Rosenband}}, \ and\ \bibinfo {author}
  {\bibfnamefont {K.-K.}\ \bibnamefont {Ni}},\ }\href {\doibase
  10.1126/science.aar7797} {\bibfield  {journal} {\bibinfo  {journal}
  {Science}\ }\textbf {\bibinfo {volume} {360}},\ \bibinfo {pages} {900}
  (\bibinfo {year} {2018})}\BibitemShut {NoStop}%
\bibitem [{\citenamefont {Weyland}\ \emph {et~al.}(2021)\citenamefont
  {Weyland}, \citenamefont {Szigeti}, \citenamefont {Hobbs}, \citenamefont
  {Ruksasakchai}, \citenamefont {Sanchez},\ and\ \citenamefont
  {Andersen}}]{Weyland2021}%
  \BibitemOpen
  \bibfield  {author} {\bibinfo {author} {\bibfnamefont {M.}~\bibnamefont
  {Weyland}}, \bibinfo {author} {\bibfnamefont {S.~S.}\ \bibnamefont
  {Szigeti}}, \bibinfo {author} {\bibfnamefont {R.~A.~B.}\ \bibnamefont
  {Hobbs}}, \bibinfo {author} {\bibfnamefont {P.}~\bibnamefont {Ruksasakchai}},
  \bibinfo {author} {\bibfnamefont {L.}~\bibnamefont {Sanchez}}, \ and\
  \bibinfo {author} {\bibfnamefont {M.~F.}\ \bibnamefont {Andersen}},\ }\href
  {\doibase 10.1103/PhysRevLett.126.083401} {\bibfield  {journal} {\bibinfo
  {journal} {Phys. Rev. Lett.}\ }\textbf {\bibinfo {volume} {126}},\ \bibinfo
  {pages} {083401} (\bibinfo {year} {2021})}\BibitemShut {NoStop}%
\bibitem [{\citenamefont {Andersen}(2022)}]{Andersen2022}%
  \BibitemOpen
  \bibfield  {author} {\bibinfo {author} {\bibfnamefont {M.~F.}\ \bibnamefont
  {Andersen}},\ }\href {\doibase 10.1080/23746149.2022.2064231} {\bibfield
  {journal} {\bibinfo  {journal} {Advances in Physics: X}\ }\textbf {\bibinfo
  {volume} {7}},\ \bibinfo {pages} {2064231} (\bibinfo {year}
  {2022})}\BibitemShut {NoStop}%
\bibitem [{\citenamefont {Weitenberg}\ \emph {et~al.}(2011)\citenamefont
  {Weitenberg}, \citenamefont {Kuhr}, \citenamefont {M\o{}lmer},\ and\
  \citenamefont {Sherson}}]{Weitenberg2011}%
  \BibitemOpen
  \bibfield  {author} {\bibinfo {author} {\bibfnamefont {C.}~\bibnamefont
  {Weitenberg}}, \bibinfo {author} {\bibfnamefont {S.}~\bibnamefont {Kuhr}},
  \bibinfo {author} {\bibfnamefont {K.}~\bibnamefont {M\o{}lmer}}, \ and\
  \bibinfo {author} {\bibfnamefont {J.~F.}\ \bibnamefont {Sherson}},\ }\href
  {\doibase 10.1103/PhysRevA.84.032322} {\bibfield  {journal} {\bibinfo
  {journal} {Phys. Rev. A}\ }\textbf {\bibinfo {volume} {84}},\ \bibinfo
  {pages} {032322} (\bibinfo {year} {2011})}\BibitemShut {NoStop}%
\bibitem [{\citenamefont {Graham}\ \emph {et~al.}(2022)\citenamefont {Graham},
  \citenamefont {Song}, \citenamefont {Scott}, \citenamefont {Poole},
  \citenamefont {Phuttitarn}, \citenamefont {Jooya}, \citenamefont {Eichler},
  \citenamefont {Jiang}, \citenamefont {Marra}, \citenamefont {Grinkemeyer},
  \citenamefont {Kwon}, \citenamefont {Ebert}, \citenamefont {Cherek},
  \citenamefont {Lichtman}, \citenamefont {Gillette}, \citenamefont {Gilbert},
  \citenamefont {Bowman}, \citenamefont {Ballance}, \citenamefont {Campbell},
  \citenamefont {Dahl}, \citenamefont {Crawford}, \citenamefont {Blunt},
  \citenamefont {Rogers}, \citenamefont {Noel},\ and\ \citenamefont
  {Saffman}}]{Graham2022}%
  \BibitemOpen
  \bibfield  {author} {\bibinfo {author} {\bibfnamefont {T.}~\bibnamefont
  {Graham}}, \bibinfo {author} {\bibfnamefont {Y.}~\bibnamefont {Song}},
  \bibinfo {author} {\bibfnamefont {J.}~\bibnamefont {Scott}}, \bibinfo
  {author} {\bibfnamefont {C.}~\bibnamefont {Poole}}, \bibinfo {author}
  {\bibfnamefont {L.}~\bibnamefont {Phuttitarn}}, \bibinfo {author}
  {\bibfnamefont {K.}~\bibnamefont {Jooya}}, \bibinfo {author} {\bibfnamefont
  {P.}~\bibnamefont {Eichler}}, \bibinfo {author} {\bibfnamefont
  {X.}~\bibnamefont {Jiang}}, \bibinfo {author} {\bibfnamefont
  {A.}~\bibnamefont {Marra}}, \bibinfo {author} {\bibfnamefont
  {B.}~\bibnamefont {Grinkemeyer}}, \bibinfo {author} {\bibfnamefont
  {M.}~\bibnamefont {Kwon}}, \bibinfo {author} {\bibfnamefont {M.}~\bibnamefont
  {Ebert}}, \bibinfo {author} {\bibfnamefont {J.}~\bibnamefont {Cherek}},
  \bibinfo {author} {\bibfnamefont {M.~T.}\ \bibnamefont {Lichtman}}, \bibinfo
  {author} {\bibfnamefont {M.}~\bibnamefont {Gillette}}, \bibinfo {author}
  {\bibfnamefont {J.}~\bibnamefont {Gilbert}}, \bibinfo {author} {\bibfnamefont
  {D.}~\bibnamefont {Bowman}}, \bibinfo {author} {\bibfnamefont
  {T.}~\bibnamefont {Ballance}}, \bibinfo {author} {\bibfnamefont
  {C.}~\bibnamefont {Campbell}}, \bibinfo {author} {\bibfnamefont {E.~D.}\
  \bibnamefont {Dahl}}, \bibinfo {author} {\bibfnamefont {O.}~\bibnamefont
  {Crawford}}, \bibinfo {author} {\bibfnamefont {N.~S.}\ \bibnamefont {Blunt}},
  \bibinfo {author} {\bibfnamefont {B.}~\bibnamefont {Rogers}}, \bibinfo
  {author} {\bibfnamefont {T.}~\bibnamefont {Noel}}, \ and\ \bibinfo {author}
  {\bibfnamefont {M.}~\bibnamefont {Saffman}},\ }\href {\doibase
  10.1038/s41586-022-04603-6} {\bibfield  {journal} {\bibinfo  {journal}
  {Nature}\ }\textbf {\bibinfo {volume} {604}},\ \bibinfo {pages} {457}
  (\bibinfo {year} {2022})}\BibitemShut {NoStop}%
\bibitem [{\citenamefont {Bernien}\ \emph {et~al.}(2017)\citenamefont
  {Bernien}, \citenamefont {Schwartz}, \citenamefont {Keesling}, \citenamefont
  {Levine}, \citenamefont {Omran}, \citenamefont {Pichler}, \citenamefont
  {Choi}, \citenamefont {Zibrov}, \citenamefont {Endres}, \citenamefont
  {Greiner}, \citenamefont {Vuletić},\ and\ \citenamefont
  {Lukin}}]{Bernien2017}%
  \BibitemOpen
  \bibfield  {author} {\bibinfo {author} {\bibfnamefont {H.}~\bibnamefont
  {Bernien}}, \bibinfo {author} {\bibfnamefont {S.}~\bibnamefont {Schwartz}},
  \bibinfo {author} {\bibfnamefont {A.}~\bibnamefont {Keesling}}, \bibinfo
  {author} {\bibfnamefont {H.}~\bibnamefont {Levine}}, \bibinfo {author}
  {\bibfnamefont {A.}~\bibnamefont {Omran}}, \bibinfo {author} {\bibfnamefont
  {H.}~\bibnamefont {Pichler}}, \bibinfo {author} {\bibfnamefont
  {S.}~\bibnamefont {Choi}}, \bibinfo {author} {\bibfnamefont {A.~S.}\
  \bibnamefont {Zibrov}}, \bibinfo {author} {\bibfnamefont {M.}~\bibnamefont
  {Endres}}, \bibinfo {author} {\bibfnamefont {M.}~\bibnamefont {Greiner}},
  \bibinfo {author} {\bibfnamefont {V.}~\bibnamefont {Vuletić}}, \ and\
  \bibinfo {author} {\bibfnamefont {M.~D.}\ \bibnamefont {Lukin}},\ }\href@noop
  {} {\bibfield  {journal} {\bibinfo  {journal} {Nature}\ }\textbf {\bibinfo
  {volume} {551}},\ \bibinfo {pages} {579} (\bibinfo {year}
  {2017})}\BibitemShut {NoStop}%
\bibitem [{\citenamefont {Browaeys}\ and\ \citenamefont
  {Lahaye}(2020)}]{Browaeys2020}%
  \BibitemOpen
  \bibfield  {author} {\bibinfo {author} {\bibfnamefont {A.}~\bibnamefont
  {Browaeys}}\ and\ \bibinfo {author} {\bibfnamefont {T.}~\bibnamefont
  {Lahaye}},\ }\href@noop {} {\bibfield  {journal} {\bibinfo  {journal} {Nature
  Physics}\ }\textbf {\bibinfo {volume} {16}},\ \bibinfo {pages} {132}
  (\bibinfo {year} {2020})}\BibitemShut {NoStop}%
\bibitem [{\citenamefont {Ðorđević}\ \emph {et~al.}(2021)\citenamefont
  {Ðorđević}, \citenamefont {Samutpraphoot}, \citenamefont {Ocola},
  \citenamefont {Bernien}, \citenamefont {Grinkemeyer}, \citenamefont
  {Dimitrova}, \citenamefont {Vuletić},\ and\ \citenamefont
  {Lukin}}]{Dordevic2021}%
  \BibitemOpen
  \bibfield  {author} {\bibinfo {author} {\bibfnamefont {T.}~\bibnamefont
  {Ðorđević}}, \bibinfo {author} {\bibfnamefont {P.}~\bibnamefont
  {Samutpraphoot}}, \bibinfo {author} {\bibfnamefont {P.~L.}\ \bibnamefont
  {Ocola}}, \bibinfo {author} {\bibfnamefont {H.}~\bibnamefont {Bernien}},
  \bibinfo {author} {\bibfnamefont {B.}~\bibnamefont {Grinkemeyer}}, \bibinfo
  {author} {\bibfnamefont {I.}~\bibnamefont {Dimitrova}}, \bibinfo {author}
  {\bibfnamefont {V.}~\bibnamefont {Vuletić}}, \ and\ \bibinfo {author}
  {\bibfnamefont {M.~D.}\ \bibnamefont {Lukin}},\ }\href {\doibase
  10.1126/science.abi9917} {\bibfield  {journal} {\bibinfo  {journal}
  {Science}\ }\textbf {\bibinfo {volume} {373}},\ \bibinfo {pages} {1511}
  (\bibinfo {year} {2021})}\BibitemShut {NoStop}%
\bibitem [{\citenamefont {Wilk}\ \emph {et~al.}(2010)\citenamefont {Wilk},
  \citenamefont {Ga\"etan}, \citenamefont {Evellin}, \citenamefont {Wolters},
  \citenamefont {Miroshnychenko}, \citenamefont {Grangier},\ and\ \citenamefont
  {Browaeys}}]{Wilk2010}%
  \BibitemOpen
  \bibfield  {author} {\bibinfo {author} {\bibfnamefont {T.}~\bibnamefont
  {Wilk}}, \bibinfo {author} {\bibfnamefont {A.}~\bibnamefont {Ga\"etan}},
  \bibinfo {author} {\bibfnamefont {C.}~\bibnamefont {Evellin}}, \bibinfo
  {author} {\bibfnamefont {J.}~\bibnamefont {Wolters}}, \bibinfo {author}
  {\bibfnamefont {Y.}~\bibnamefont {Miroshnychenko}}, \bibinfo {author}
  {\bibfnamefont {P.}~\bibnamefont {Grangier}}, \ and\ \bibinfo {author}
  {\bibfnamefont {A.}~\bibnamefont {Browaeys}},\ }\href {\doibase
  10.1103/PhysRevLett.104.010502} {\bibfield  {journal} {\bibinfo  {journal}
  {Phys. Rev. Lett.}\ }\textbf {\bibinfo {volume} {104}},\ \bibinfo {pages}
  {010502} (\bibinfo {year} {2010})}\BibitemShut {NoStop}%
\bibitem [{\citenamefont {Fuhrmanek}\ \emph {et~al.}(2010)\citenamefont
  {Fuhrmanek}, \citenamefont {Sortais}, \citenamefont {Grangier},\ and\
  \citenamefont {Browaeys}}]{Fuhrmanek2010}%
  \BibitemOpen
  \bibfield  {author} {\bibinfo {author} {\bibfnamefont {A.}~\bibnamefont
  {Fuhrmanek}}, \bibinfo {author} {\bibfnamefont {Y.~R.~P.}\ \bibnamefont
  {Sortais}}, \bibinfo {author} {\bibfnamefont {P.}~\bibnamefont {Grangier}}, \
  and\ \bibinfo {author} {\bibfnamefont {A.}~\bibnamefont {Browaeys}},\ }\href
  {\doibase 10.1103/PhysRevA.82.023623} {\bibfield  {journal} {\bibinfo
  {journal} {Phys. Rev. A}\ }\textbf {\bibinfo {volume} {82}},\ \bibinfo
  {pages} {023623} (\bibinfo {year} {2010})}\BibitemShut {NoStop}%
\bibitem [{\citenamefont {Reynolds}\ \emph {et~al.}(2020)\citenamefont
  {Reynolds}, \citenamefont {Schwartz}, \citenamefont {Ebling}, \citenamefont
  {Weyland}, \citenamefont {Brand},\ and\ \citenamefont
  {Andersen}}]{Reynolds2020}%
  \BibitemOpen
  \bibfield  {author} {\bibinfo {author} {\bibfnamefont {L.~A.}\ \bibnamefont
  {Reynolds}}, \bibinfo {author} {\bibfnamefont {E.}~\bibnamefont {Schwartz}},
  \bibinfo {author} {\bibfnamefont {U.}~\bibnamefont {Ebling}}, \bibinfo
  {author} {\bibfnamefont {M.}~\bibnamefont {Weyland}}, \bibinfo {author}
  {\bibfnamefont {J.}~\bibnamefont {Brand}}, \ and\ \bibinfo {author}
  {\bibfnamefont {M.~F.}\ \bibnamefont {Andersen}},\ }\href {\doibase
  10.1103/PhysRevLett.124.073401} {\bibfield  {journal} {\bibinfo  {journal}
  {Phys. Rev. Lett.}\ }\textbf {\bibinfo {volume} {124}},\ \bibinfo {pages}
  {073401} (\bibinfo {year} {2020})}\BibitemShut {NoStop}%
\bibitem [{\citenamefont {Gr\"un}\ \emph {et~al.}(2024)\citenamefont {Gr\"un},
  \citenamefont {White}, \citenamefont {Ortu}, \citenamefont {Di~Carli},
  \citenamefont {Edri}, \citenamefont {Lepers}, \citenamefont {Mark},\ and\
  \citenamefont {Ferlaino}}]{Gruen2024}%
  \BibitemOpen
  \bibfield  {author} {\bibinfo {author} {\bibfnamefont {D.~S.}\ \bibnamefont
  {Gr\"un}}, \bibinfo {author} {\bibfnamefont {S.~J.~M.}\ \bibnamefont
  {White}}, \bibinfo {author} {\bibfnamefont {A.}~\bibnamefont {Ortu}},
  \bibinfo {author} {\bibfnamefont {A.}~\bibnamefont {Di~Carli}}, \bibinfo
  {author} {\bibfnamefont {H.}~\bibnamefont {Edri}}, \bibinfo {author}
  {\bibfnamefont {M.}~\bibnamefont {Lepers}}, \bibinfo {author} {\bibfnamefont
  {M.~J.}\ \bibnamefont {Mark}}, \ and\ \bibinfo {author} {\bibfnamefont
  {F.}~\bibnamefont {Ferlaino}},\ }\href {\doibase
  10.1103/PhysRevLett.133.223402} {\bibfield  {journal} {\bibinfo  {journal}
  {Phys. Rev. Lett.}\ }\textbf {\bibinfo {volume} {133}},\ \bibinfo {pages}
  {223402} (\bibinfo {year} {2024})}\BibitemShut {NoStop}%
\bibitem [{\citenamefont {Zhang}\ \emph {et~al.}(2020)\citenamefont {Zhang},
  \citenamefont {Yu}, \citenamefont {Cairncross}, \citenamefont {Wang},
  \citenamefont {Picard}, \citenamefont {Hood}, \citenamefont {Lin},
  \citenamefont {Hutson},\ and\ \citenamefont {Ni}}]{Zhang2020}%
  \BibitemOpen
  \bibfield  {author} {\bibinfo {author} {\bibfnamefont {J.~T.}\ \bibnamefont
  {Zhang}}, \bibinfo {author} {\bibfnamefont {Y.}~\bibnamefont {Yu}}, \bibinfo
  {author} {\bibfnamefont {W.~B.}\ \bibnamefont {Cairncross}}, \bibinfo
  {author} {\bibfnamefont {K.}~\bibnamefont {Wang}}, \bibinfo {author}
  {\bibfnamefont {L.~R.~B.}\ \bibnamefont {Picard}}, \bibinfo {author}
  {\bibfnamefont {J.~D.}\ \bibnamefont {Hood}}, \bibinfo {author}
  {\bibfnamefont {Y.-W.}\ \bibnamefont {Lin}}, \bibinfo {author} {\bibfnamefont
  {J.~M.}\ \bibnamefont {Hutson}}, \ and\ \bibinfo {author} {\bibfnamefont
  {K.-K.}\ \bibnamefont {Ni}},\ }\href {\doibase
  10.1103/PhysRevLett.124.253401} {\bibfield  {journal} {\bibinfo  {journal}
  {Phys. Rev. Lett.}\ }\textbf {\bibinfo {volume} {124}},\ \bibinfo {pages}
  {253401} (\bibinfo {year} {2020})}\BibitemShut {NoStop}%
\bibitem [{\citenamefont {Kuhr}\ \emph {et~al.}(2005)\citenamefont {Kuhr},
  \citenamefont {Alt}, \citenamefont {Schrader}, \citenamefont {Dotsenko},
  \citenamefont {Miroshnychenko}, \citenamefont {Rauschenbeutel},\ and\
  \citenamefont {Meschede}}]{Kuhr2005}%
  \BibitemOpen
  \bibfield  {author} {\bibinfo {author} {\bibfnamefont {S.}~\bibnamefont
  {Kuhr}}, \bibinfo {author} {\bibfnamefont {W.}~\bibnamefont {Alt}}, \bibinfo
  {author} {\bibfnamefont {D.}~\bibnamefont {Schrader}}, \bibinfo {author}
  {\bibfnamefont {I.}~\bibnamefont {Dotsenko}}, \bibinfo {author}
  {\bibfnamefont {Y.}~\bibnamefont {Miroshnychenko}}, \bibinfo {author}
  {\bibfnamefont {A.}~\bibnamefont {Rauschenbeutel}}, \ and\ \bibinfo {author}
  {\bibfnamefont {D.}~\bibnamefont {Meschede}},\ }\href {\doibase
  10.1103/PhysRevA.72.023406} {\bibfield  {journal} {\bibinfo  {journal} {Phys.
  Rev. A}\ }\textbf {\bibinfo {volume} {72}},\ \bibinfo {pages} {023406}
  (\bibinfo {year} {2005})}\BibitemShut {NoStop}%
\bibitem [{\citenamefont {Xu}\ \emph {et~al.}(2021)\citenamefont {Xu},
  \citenamefont {Venkatramani}, \citenamefont {Cantu}, \citenamefont
  {Šumarac}, \citenamefont {Klüsener}, \citenamefont {Lukin},\ and\
  \citenamefont {Vuletić}}]{Xu2021}%
  \BibitemOpen
  \bibfield  {author} {\bibinfo {author} {\bibfnamefont {W.}~\bibnamefont
  {Xu}}, \bibinfo {author} {\bibfnamefont {A.~V.}\ \bibnamefont
  {Venkatramani}}, \bibinfo {author} {\bibfnamefont {S.~H.}\ \bibnamefont
  {Cantu}}, \bibinfo {author} {\bibfnamefont {T.}~\bibnamefont {Šumarac}},
  \bibinfo {author} {\bibfnamefont {V.}~\bibnamefont {Klüsener}}, \bibinfo
  {author} {\bibfnamefont {M.~D.}\ \bibnamefont {Lukin}}, \ and\ \bibinfo
  {author} {\bibfnamefont {V.}~\bibnamefont {Vuletić}},\ }\href@noop {}
  {\bibfield  {journal} {\bibinfo  {journal} {Phys. Rev. Lett.}\ }\textbf
  {\bibinfo {volume} {127}},\ \bibinfo {pages} {050501} (\bibinfo {year}
  {2021})}\BibitemShut {NoStop}%
\bibitem [{\citenamefont {Bloch}\ \emph {et~al.}(2023)\citenamefont {Bloch},
  \citenamefont {Hofer}, \citenamefont {Cohen}, \citenamefont {Browaeys},\ and\
  \citenamefont {Ferrier-Barbut}}]{Bloch2023}%
  \BibitemOpen
  \bibfield  {author} {\bibinfo {author} {\bibfnamefont {D.}~\bibnamefont
  {Bloch}}, \bibinfo {author} {\bibfnamefont {B.}~\bibnamefont {Hofer}},
  \bibinfo {author} {\bibfnamefont {S.~R.}\ \bibnamefont {Cohen}}, \bibinfo
  {author} {\bibfnamefont {A.}~\bibnamefont {Browaeys}}, \ and\ \bibinfo
  {author} {\bibfnamefont {I.}~\bibnamefont {Ferrier-Barbut}},\ }\href
  {\doibase 10.1103/PhysRevLett.131.203401} {\bibfield  {journal} {\bibinfo
  {journal} {Phys. Rev. Lett.}\ }\textbf {\bibinfo {volume} {131}},\ \bibinfo
  {pages} {203401} (\bibinfo {year} {2023})}\BibitemShut {NoStop}%
\bibitem [{\citenamefont {McGovern}\ \emph {et~al.}(2011)\citenamefont
  {McGovern}, \citenamefont {Hilliard}, \citenamefont {Gr\"{u}nzweig},\ and\
  \citenamefont {Andersen}}]{McGovern2011}%
  \BibitemOpen
  \bibfield  {author} {\bibinfo {author} {\bibfnamefont {M.}~\bibnamefont
  {McGovern}}, \bibinfo {author} {\bibfnamefont {A.~J.}\ \bibnamefont
  {Hilliard}}, \bibinfo {author} {\bibfnamefont {T.}~\bibnamefont
  {Gr\"{u}nzweig}}, \ and\ \bibinfo {author} {\bibfnamefont {M.~F.}\
  \bibnamefont {Andersen}},\ }\href {\doibase 10.1364/OL.36.001041} {\bibfield
  {journal} {\bibinfo  {journal} {Opt. Lett.}\ }\textbf {\bibinfo {volume}
  {36}},\ \bibinfo {pages} {1041} (\bibinfo {year} {2011})}\BibitemShut
  {NoStop}%
\bibitem [{\citenamefont {Meng}\ \emph {et~al.}(2020)\citenamefont {Meng},
  \citenamefont {Liedl}, \citenamefont {Pucher}, \citenamefont
  {Rauschenbeutel},\ and\ \citenamefont {Schneeweiss}}]{Meng2020}%
  \BibitemOpen
  \bibfield  {author} {\bibinfo {author} {\bibfnamefont {Y.}~\bibnamefont
  {Meng}}, \bibinfo {author} {\bibfnamefont {C.}~\bibnamefont {Liedl}},
  \bibinfo {author} {\bibfnamefont {S.}~\bibnamefont {Pucher}}, \bibinfo
  {author} {\bibfnamefont {A.}~\bibnamefont {Rauschenbeutel}}, \ and\ \bibinfo
  {author} {\bibfnamefont {P.}~\bibnamefont {Schneeweiss}},\ }\href {\doibase
  10.1103/PhysRevLett.125.053603} {\bibfield  {journal} {\bibinfo  {journal}
  {Phys. Rev. Lett.}\ }\textbf {\bibinfo {volume} {125}},\ \bibinfo {pages}
  {053603} (\bibinfo {year} {2020})}\BibitemShut {NoStop}%
\bibitem [{\citenamefont {Sompet}\ \emph {et~al.}(2019)\citenamefont {Sompet},
  \citenamefont {Szigeti}, \citenamefont {Schwartz}, \citenamefont {Bradley},\
  and\ \citenamefont {Andersen}}]{Sompet2019}%
  \BibitemOpen
  \bibfield  {author} {\bibinfo {author} {\bibfnamefont {P.}~\bibnamefont
  {Sompet}}, \bibinfo {author} {\bibfnamefont {S.~S.}\ \bibnamefont {Szigeti}},
  \bibinfo {author} {\bibfnamefont {E.}~\bibnamefont {Schwartz}}, \bibinfo
  {author} {\bibfnamefont {A.~S.}\ \bibnamefont {Bradley}}, \ and\ \bibinfo
  {author} {\bibfnamefont {M.~F.}\ \bibnamefont {Andersen}},\ }\href@noop {}
  {\bibfield  {journal} {\bibinfo  {journal} {Nature communications}\ }\textbf
  {\bibinfo {volume} {10}},\ \bibinfo {pages} {1} (\bibinfo {year}
  {2019})}\BibitemShut {NoStop}%
\bibitem [{\citenamefont {Carpentier}\ \emph {et~al.}(2013)\citenamefont
  {Carpentier}, \citenamefont {Fung}, \citenamefont {Sompet}, \citenamefont
  {Hilliard}, \citenamefont {Walker},\ and\ \citenamefont
  {Andersen}}]{Carpentier2013}%
  \BibitemOpen
  \bibfield  {author} {\bibinfo {author} {\bibfnamefont {A.~V.}\ \bibnamefont
  {Carpentier}}, \bibinfo {author} {\bibfnamefont {Y.~H.}\ \bibnamefont
  {Fung}}, \bibinfo {author} {\bibfnamefont {P.}~\bibnamefont {Sompet}},
  \bibinfo {author} {\bibfnamefont {A.~J.}\ \bibnamefont {Hilliard}}, \bibinfo
  {author} {\bibfnamefont {T.~G.}\ \bibnamefont {Walker}}, \ and\ \bibinfo
  {author} {\bibfnamefont {M.~F.}\ \bibnamefont {Andersen}},\ }\href@noop {}
  {\bibfield  {journal} {\bibinfo  {journal} {Laser Phys. Lett.}\ }\textbf
  {\bibinfo {volume} {10}},\ \bibinfo {pages} {125501} (\bibinfo {year}
  {2013})}\BibitemShut {NoStop}%
\bibitem [{\citenamefont {Hilliard}\ \emph {et~al.}(2015)\citenamefont
  {Hilliard}, \citenamefont {Fung}, \citenamefont {Sompet}, \citenamefont
  {Carpentier},\ and\ \citenamefont {Andersen}}]{Hilliard2015}%
  \BibitemOpen
  \bibfield  {author} {\bibinfo {author} {\bibfnamefont {A.~J.}\ \bibnamefont
  {Hilliard}}, \bibinfo {author} {\bibfnamefont {Y.~H.}\ \bibnamefont {Fung}},
  \bibinfo {author} {\bibfnamefont {P.}~\bibnamefont {Sompet}}, \bibinfo
  {author} {\bibfnamefont {A.~V.}\ \bibnamefont {Carpentier}}, \ and\ \bibinfo
  {author} {\bibfnamefont {M.~F.}\ \bibnamefont {Andersen}},\ }\href {\doibase
  10.1103/PhysRevA.91.053414} {\bibfield  {journal} {\bibinfo  {journal} {Phys.
  Rev. A}\ }\textbf {\bibinfo {volume} {91}},\ \bibinfo {pages} {053414}
  (\bibinfo {year} {2015})}\BibitemShut {NoStop}%
\bibitem [{\citenamefont {Fung}\ \emph {et~al.}(2016)\citenamefont {Fung},
  \citenamefont {Sompet},\ and\ \citenamefont {Andersen}}]{Fung2016}%
  \BibitemOpen
  \bibfield  {author} {\bibinfo {author} {\bibfnamefont {Y.~H.}\ \bibnamefont
  {Fung}}, \bibinfo {author} {\bibfnamefont {P.}~\bibnamefont {Sompet}}, \ and\
  \bibinfo {author} {\bibfnamefont {M.~F.}\ \bibnamefont {Andersen}},\ }\href
  {https://www.mdpi.com/2227-7080/4/1/4} {\bibfield  {journal} {\bibinfo
  {journal} {Technologies}\ }\textbf {\bibinfo {volume} {4}} (\bibinfo {year}
  {2016})}\BibitemShut {NoStop}%
\bibitem [{\citenamefont {Fung}\ and\ \citenamefont
  {Andersen}(2015)}]{Fung2015}%
  \BibitemOpen
  \bibfield  {author} {\bibinfo {author} {\bibfnamefont {Y.~H.}\ \bibnamefont
  {Fung}}\ and\ \bibinfo {author} {\bibfnamefont {M.~F.}\ \bibnamefont
  {Andersen}},\ }\href@noop {} {\bibfield  {journal} {\bibinfo  {journal} {New
  J. Phys.}\ }\textbf {\bibinfo {volume} {17}},\ \bibinfo {pages} {073011}
  (\bibinfo {year} {2015})}\BibitemShut {NoStop}%
\bibitem [{\citenamefont {Goodman}(2015)}]{goodman2015}%
  \BibitemOpen
  \bibfield  {author} {\bibinfo {author} {\bibfnamefont {J.~W.}\ \bibnamefont
  {Goodman}},\ }\href@noop {} {\emph {\bibinfo {title} {Statistical Optics}}}\
  (\bibinfo  {publisher} {John Wiley \& Sons Inc.},\ \bibinfo {year}
  {2015})\BibitemShut {NoStop}%
\bibitem [{\citenamefont {Grünzweig}\ \emph {et~al.}(2011)\citenamefont
  {Grünzweig}, \citenamefont {McGovern}, \citenamefont {Hillard},\ and\
  \citenamefont {Andersen}}]{Gruenzweig2011}%
  \BibitemOpen
  \bibfield  {author} {\bibinfo {author} {\bibfnamefont {T.}~\bibnamefont
  {Grünzweig}}, \bibinfo {author} {\bibfnamefont {M.}~\bibnamefont
  {McGovern}}, \bibinfo {author} {\bibfnamefont {A.~J.}\ \bibnamefont
  {Hillard}}, \ and\ \bibinfo {author} {\bibfnamefont {M.~F.}\ \bibnamefont
  {Andersen}},\ }\href@noop {} {\bibfield  {journal} {\bibinfo  {journal}
  {Quantum Inf. Process.}\ }\textbf {\bibinfo {volume} {10}},\ \bibinfo {pages}
  {925} (\bibinfo {year} {2011})}\BibitemShut {NoStop}%
\bibitem [{\citenamefont {Stefany~Coxe}\ and\ \citenamefont
  {Aiken}(2009)}]{Coxe2009}%
  \BibitemOpen
  \bibfield  {author} {\bibinfo {author} {\bibfnamefont {S.~G.~W.}\
  \bibnamefont {Stefany~Coxe}}\ and\ \bibinfo {author} {\bibfnamefont {L.~S.}\
  \bibnamefont {Aiken}},\ }\href {\doibase 10.1080/00223890802634175}
  {\bibfield  {journal} {\bibinfo  {journal} {Journal of Personality
  Assessment}\ }\textbf {\bibinfo {volume} {91}},\ \bibinfo {pages} {121}
  (\bibinfo {year} {2009})}\BibitemShut {NoStop}%
\bibitem [{Note1()}]{Note1}%
  \BibitemOpen
  \bibinfo {note} {We use the MLE function of MATLAB for this.}\BibitemShut
  {Stop}%
\bibitem [{\citenamefont {Raikov}(1937)}]{Raikov1937}%
  \BibitemOpen
  \bibfield  {author} {\bibinfo {author} {\bibfnamefont {D.}~\bibnamefont
  {Raikov}},\ }\href@noop {} {\bibfield  {journal} {\bibinfo  {journal}
  {Comptes Rendus de l'Academie des Sciences de l'URSS}\ }\textbf {\bibinfo
  {volume} {14}},\ \bibinfo {pages} {9} (\bibinfo {year} {1937})}\BibitemShut
  {NoStop}%
\bibitem [{\citenamefont {Barlow}(1989)}]{Barlow1989}%
  \BibitemOpen
  \bibfield  {author} {\bibinfo {author} {\bibfnamefont {R.}~\bibnamefont
  {Barlow}},\ }\href@noop {} {\emph {\bibinfo {title} {Statistics: A Guide to
  the Use of Statistical Methods in the Physical Sciences}}}\ (\bibinfo
  {publisher} {John Wiley \& Sons Inc.},\ \bibinfo {year} {1989})\BibitemShut
  {NoStop}%
\bibitem [{\citenamefont {Taylor}(1997)}]{taylor}%
  \BibitemOpen
  \bibfield  {author} {\bibinfo {author} {\bibfnamefont {J.~R.}\ \bibnamefont
  {Taylor}},\ }\href@noop {} {\emph {\bibinfo {title} {An Introduction to Error
  Analysis}}},\ \bibinfo {edition} {2nd}\ ed.\ (\bibinfo  {publisher}
  {University Science Books},\ \bibinfo {year} {1997})\BibitemShut {NoStop}%
\bibitem [{Note2()}]{Note2}%
  \BibitemOpen
  \bibinfo {note} {While more sophisticated fitting algorithms would be able to
  perform this fit, we chose this simple approach as it can also serve as a
  robust method when directly fitting to experimentally measured control
  distributions with no functional model, as described in Sec.~\ref
  {Sec6:DataFit}.}\BibitemShut {Stop}%
\bibitem [{\citenamefont {Sompet}\ \emph {et~al.}(2013)\citenamefont {Sompet},
  \citenamefont {Carpentier}, \citenamefont {Fung}, \citenamefont {McGovern},\
  and\ \citenamefont {Andersen}}]{Sompet2013}%
  \BibitemOpen
  \bibfield  {author} {\bibinfo {author} {\bibfnamefont {P.}~\bibnamefont
  {Sompet}}, \bibinfo {author} {\bibfnamefont {A.~V.}\ \bibnamefont
  {Carpentier}}, \bibinfo {author} {\bibfnamefont {Y.~H.}\ \bibnamefont
  {Fung}}, \bibinfo {author} {\bibfnamefont {M.}~\bibnamefont {McGovern}}, \
  and\ \bibinfo {author} {\bibfnamefont {M.~F.}\ \bibnamefont {Andersen}},\
  }\href {\doibase 10.1103/PhysRevA.88.051401} {\bibfield  {journal} {\bibinfo
  {journal} {Phys. Rev. A}\ }\textbf {\bibinfo {volume} {88}},\ \bibinfo
  {pages} {051401} (\bibinfo {year} {2013})}\BibitemShut {NoStop}%
\bibitem [{\citenamefont {Gl{\"{u}}senkamp}(2020)}]{Glusenkamp2020}%
  \BibitemOpen
  \bibfield  {author} {\bibinfo {author} {\bibfnamefont {T.}~\bibnamefont
  {Gl{\"{u}}senkamp}},\ }\href@noop {} {\bibfield  {journal} {\bibinfo
  {journal} {Journal of Instrumentation}\ }\textbf {\bibinfo {volume} {15}},\
  \bibinfo {pages} {P01035} (\bibinfo {year} {2020})}\BibitemShut {NoStop}%
\bibitem [{\citenamefont {Bohm}\ and\ \citenamefont {Zech}(2012)}]{Bohm2012}%
  \BibitemOpen
  \bibfield  {author} {\bibinfo {author} {\bibfnamefont {G.}~\bibnamefont
  {Bohm}}\ and\ \bibinfo {author} {\bibfnamefont {G.}~\bibnamefont {Zech}},\
  }\href {\doibase https://doi.org/10.1016/j.nima.2012.06.021} {\bibfield
  {journal} {\bibinfo  {journal} {Nucl. Instrum. Methods Phys. Res. A: Accel.
  Spectrom. Detect. Assoc. Equip.}\ }\textbf {\bibinfo {volume} {691}},\
  \bibinfo {pages} {171} (\bibinfo {year} {2012})}\BibitemShut {NoStop}%
\bibitem [{\citenamefont {Bohm}\ and\ \citenamefont {Zech}(2014)}]{Bohm2014}%
  \BibitemOpen
  \bibfield  {author} {\bibinfo {author} {\bibfnamefont {G.}~\bibnamefont
  {Bohm}}\ and\ \bibinfo {author} {\bibfnamefont {G.}~\bibnamefont {Zech}},\
  }\href {\doibase 10.1016/j.nima.2014.02.021} {\bibfield  {journal} {\bibinfo
  {journal} {Nucl. Instrum. Methods Phys. Res. A: Accel. Spectrom. Detect.
  Assoc. Equip.}\ }\textbf {\bibinfo {volume} {748}},\ \bibinfo {pages} {1}
  (\bibinfo {year} {2014})},\ \Eprint {http://arxiv.org/abs/1309.1287}
  {arXiv:1309.1287} \BibitemShut {NoStop}%
\bibitem [{\citenamefont {Chirkin}(2013)}]{Chirkin2013}%
  \BibitemOpen
  \bibfield  {author} {\bibinfo {author} {\bibfnamefont {D.}~\bibnamefont
  {Chirkin}},\ }\href@noop {} {\bibfield  {journal} {\bibinfo  {journal}
  {arXiv:1304.0735}\ } (\bibinfo {year} {2013})}\BibitemShut {NoStop}%
\bibitem [{\citenamefont {Glüsenkamp}(2018)}]{Glusenkamp2018}%
  \BibitemOpen
  \bibfield  {author} {\bibinfo {author} {\bibfnamefont {T.}~\bibnamefont
  {Glüsenkamp}},\ }\href {\doibase 10.1140/epjp/i2018-12042-x} {\bibfield
  {journal} {\bibinfo  {journal} {Eur. Phys. J. Plus}\ }\textbf {\bibinfo
  {volume} {133}},\ \bibinfo {pages} {218} (\bibinfo {year}
  {2018})}\BibitemShut {NoStop}%
\bibitem [{\citenamefont {Arg{\"{u}}elles}\ \emph {et~al.}(2019)\citenamefont
  {Arg{\"{u}}elles}, \citenamefont {Schneider},\ and\ \citenamefont
  {Yuan}}]{Arguelles2019}%
  \BibitemOpen
  \bibfield  {author} {\bibinfo {author} {\bibfnamefont {C.~A.}\ \bibnamefont
  {Arg{\"{u}}elles}}, \bibinfo {author} {\bibfnamefont {A.}~\bibnamefont
  {Schneider}}, \ and\ \bibinfo {author} {\bibfnamefont {T.}~\bibnamefont
  {Yuan}},\ }\href@noop {} {\bibfield  {journal} {\bibinfo  {journal} {Journal
  of High Energy Physics}\ }\textbf {\bibinfo {volume} {2019}},\ \bibinfo
  {pages} {30} (\bibinfo {year} {2019})}\BibitemShut {NoStop}%
\bibitem [{Note3()}]{Note3}%
  \BibitemOpen
  \bibinfo {note} {When the fit fails in a small region of $\theta $-space, the
  average error increases without a significant impact on the $1\sigma $
  confidence interval. As a result the average error starts being located
  towards the upped edge or even outside of the confidence
  interval.}\BibitemShut {Stop}%
\bibitem [{\citenamefont {Chernick}(2007)}]{Chernick}%
  \BibitemOpen
  \bibfield  {author} {\bibinfo {author} {\bibfnamefont {M.~R.}\ \bibnamefont
  {Chernick}},\ }\href@noop {} {\emph {\bibinfo {title} {Bootstrap Methods: A
  Guide for Practitioners and Researchers}}}\ (\bibinfo  {publisher} {John
  Wiley \& Sons Inc.},\ \bibinfo {year} {2007})\BibitemShut {NoStop}%
\end{thebibliography}%

\end{document}